\documentclass[12pt]{article}

\usepackage{graphpap}
\usepackage{graphics}

\newcommand{\calO}{{\cal O}}
\newcommand{\Obar}{\bar{\calO}}
\newcommand{\I}{{\cal I}}
\newcommand{\Idag}{\I^\dag}
\newcommand{\J}{{\cal J}}
\newcommand{\Jdag}{\J^\dag}
\newcommand{\LLDA}{{\cal L}_{LDA}}
\newcommand{\ELDA}{E_{LDA}}
\newcommand{\psiset}{\{\psi_i\}}
\newcommand{\psirset}{\{\psi_i(r)\}}
\newcommand{\exc}{\epsilon_{xc}}
\newcommand{\excprime}{\epsilon'_{xc}}
\newcommand{\abi}{{\em ab initio} }

\newcommand{\codeindent}{\hspace*{9ex}}

\begin{document}

\title{New Algebraic Formulation of Density Functional
Calculation}

\author{Sohrab Ismail-Beigi$^\dag$ and T.A.~Arias$^\ddag$\\
\\
$\dag$ Department of Physics, Massachusetts Institute of Technology,\\
Cambridge, Massachusetts 02139\\
\\
$\ddag$ Laboratory of Atomic and Solid State Physics, Cornell
University,\\
Ithaca, New York 14853 \\
and\\
$\ddag$ Research Laboratory of Electronics, Massachusetts Institute
of Technology,\\
Cambridge, Massachusetts 02139\\
\\
Suggested PACS codes: 71.15.-m, 71.15.Ap, 71.15.Fv, 71.15.Hx.
structure methods, electronic structure calculations, high\\
performance computing, parallel computing, computer languages}

\date{In press: Comp. Phys. Comm., vol. {\bf 128}, pp. 1-45 (June
2000).}  \maketitle

\begin{abstract}
This article addresses a fundamental problem faced by the community
employing single-particle \abi methods: the lack of an effective
formalism for the rapid exploration and exchange of new methods.  To
rectify this, we introduce a new, basis-set independent, matrix-based
formulation of generalized density functional theories which reduces
the development, implementation, and dissemination of new techniques
to the derivation and transcription of a few lines of algebra.  This
new framework enables us to concisely demystify the inner workings of
fully functional, highly efficient modern \abi codes and to give
complete instructions for their construction for calculations
employing arbitrary basis sets.  Within this framework, we also
discuss in full detail a variety of leading-edge techniques,
minimization algorithms, and highly efficient computational kernels
for use with scalar as well as shared and distributed-memory
supercomputer architectures.
\end{abstract}

\tableofcontents
\listoftables
\listoffigures

\maketitle
\vspace{.5cm}

\section{Introduction}

This work gives a self-contained description of how to build a highly
flexible, portable density-functional production code which attains
significant fractions of peak performance on scalar cached
architectures, shared-memory processors (SMP), and distributed-memory
processors (DMP).  More importantly, however, this work introduces a
new formalism, DFT++, for the development, implementation, and
dissemination of new \abi generalized functional theoretic techniques
among researchers.  The most well-known and widely used generalized
functional theory (GFT) is density-functional theory, where the energy
of the system is parametrized as a functional of the electron density.
Although the formalism presented here is applicable to other
single-particle GFTs, such as self-interaction correction or
Hartree-Fock theory, for concreteness we concentrate primarily on
density functional theory (DFT).

This formalism is of particular interest to those on the forefront of
exploring new \abi techniques and novel applications of such in the
physical sciences.  It allows practitioners to quickly introduce new
physics and techniques without expenditure of significant effort in
debugging and optimizing or in developing entirely new software
packages.  It does so by providing a new, compact, and explicit
matrix-based language for expressing GFT calculations, which allows
new codes to be ``derived'' through straightforward formal
manipulations.  It also provides a high degree of modularity, a great
aid in maintaining high computational performance.

This language may be thought of as being for GFT what the Dirac
notation is for quantum mechanics: a fully explicit notation free of
burdensome details which permits the ready performance of complex
manipulations with focus on physical content.  Direct application of
the Dirac notation to GFT is particularly cumbersome because in
single-particle theories, the quantum state of the system is not
represented by a single ket but rather a collection of kets,
necessitating a great deal of indexing.  Previous attempts to work
with the Dirac notation while eliminating this indexing have included
construction of column vectors whose entries were kets \cite{ACprl}
but such constructions have proved awkward because, ultimately, kets
are members of an abstract Hilbert space and are not the fundamental
objects of an actual calculation.

The foundation of the new DFT++ formalism is the observation that all
the necessary computations in an \abi calculation can be expressed
explicitly as standard linear-algebraic operations upon the actual
computational representation of the quantum state without reference to
complicated indexing or to the underlying basis set.  With traditional
approaches, differentiating the energy functional, which is required
for self-consistent solution for the single-particle orbitals, is a
frequent source of difficulty.  Issues arise such as the distinction
between wave functions and their duals, covariant versus contravariant
quantities, establishing a consistent set of normalization
conventions, and translation from continuum functional derivatives to
their discrete computational representations.  However, by expressing
the energy explicitly in our formalism, all these difficulties are
automatically avoided by straightforward differentiation of a
well-defined linear-algebraic expression.

This new formalism allows not only for ease of formal manipulations
but also for direct transcription of the resulting expressions into
software, i.e. literal typing of physical expressions in their matrix
form into lines of computer code.  Literal transcription of operations
such as matrix addition and multiplication is possible through the use
of any of the modern, high-level computer languages which allows for
the definition of new object types (e.g. vectors and matrices) {\em
and} the action of the standard operators such as ``+'', ``-'', or
``*'' upon them.  Once the basic operators have been implemented, the
task of developing and debugging is simplified to checking the
formulae which have been entered into the software.  This allows the
researcher to modify or extend the software and explore entirely new
physical ideas rapidly.  Finally, a very important practical benefit
of using matrix operations wherever possible is that the theory of
attaining peak performance on modern computers is well developed for
matrix-matrix multiplication.

The high level of modularity which naturally emerges within the DFT++
formalism compartmentalizes and isolates from one another the primary
areas of research in electronic structure calculation: (i) derivation
of new physical approaches, (ii) development of effective numerical
techniques for reaching self-consistency, and (iii) optimization and
parallelization of the underlying computational kernels.  This
compartmentalization brings the significant advantage that researchers
with specialized skills can explore effectively the areas which
pertain to them, without concerning themselves with the areas with
which they are less familiar.

A few anecdotes from our own experience serve to illustrate the
efficacy of this approach.  The extension of our production software
to include electron spin through the local spin-density approximation
(LSDA) required a student with no prior familiarity with our software
only one-half week to gain that familiarity, three days to redefine
the software objects to include spin, and less than one day to
implement and debug the new physics.  The time it took another student
to develop, implement, explore, and fine-tune the new numerical
technique of Section~\ref{sec:subspacerot} was less than a week.
Finally, our experience with parallelization and optimization has been
similarly successful.  To parallelize our software for use with an SMP
(using threads) required a student starting with no prior knowledge of
parallelization two weeks to develop a code which {\em sustains} an
average per processor FLOP rate of 80\% of the processor clock speed.
(See Section \ref{sec:SMPparallel} for details.)  Finally, for
massively parallel applications, the development of an efficient DMP
code (based on MPI), a task which often requires a year or more,
required two students working together approximately two months to
complete.

\begin{figure}[t!b!]
\begin{center}
\resizebox{5.5in}{!}{\includegraphics{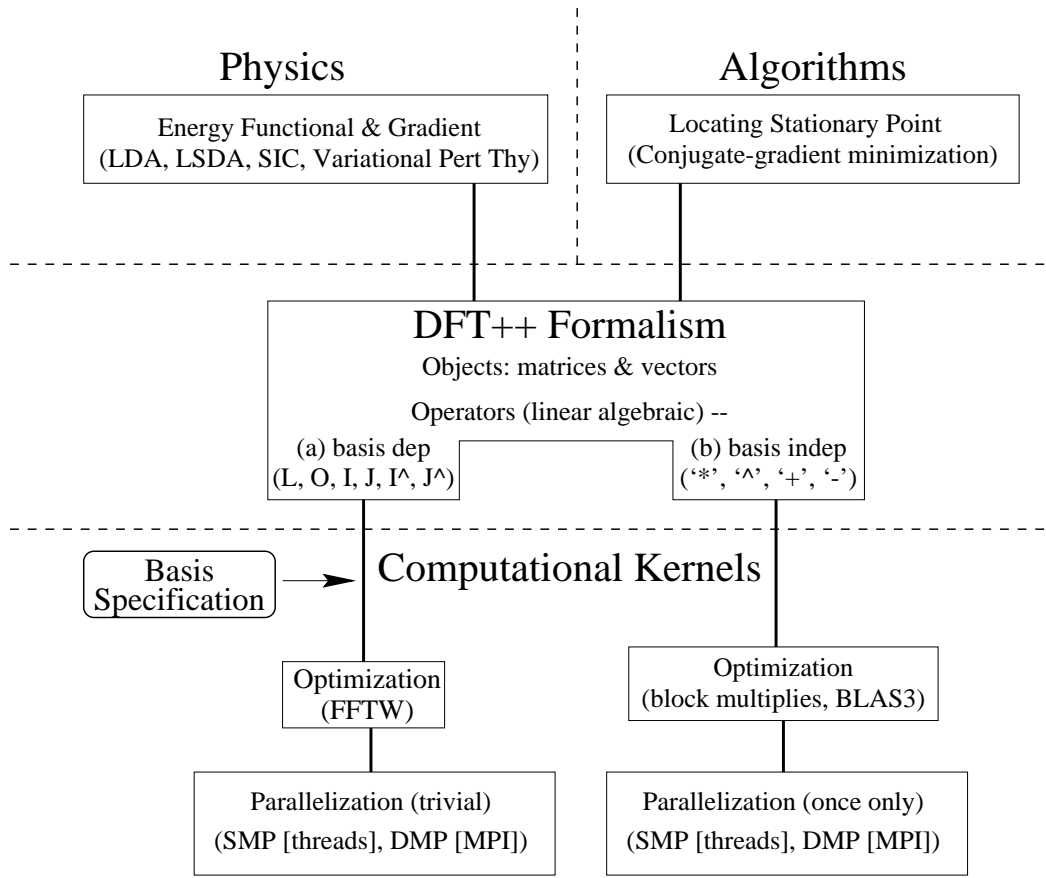}}
\caption{\label{fig:overview}Overview of DFT++ formalism}
\end{center}
\end{figure}

\section{Overview}

Figure \ref{fig:overview} both illustrates the interconnections among
the primary areas of active research in modern electronic structure
calculations and serves as a road-map for the content of this article.
The figure emphasizes how the DFT++ formalism forms an effective
central bridge connecting these areas.

Reduction to practice of new physical approaches generally requires
expressions for an energy functional and the derivatives of that
functional, as indicated in the upper-left portion of the figure.  Our
discussion begins in Section~\ref{sec:lagr} with an exposition of the
mathematical framework which we employ throughout this work, a
Lagrangian formulation of generalized density functional theories.  In
Section~\ref{sec:matrixnotation} we introduce our matrix-based
formalism using density-functional theory (DFT) within the
local-density approximation (LDA) \cite{KohnSham} as a case study,
deriving the requisite expressions for the energy functional and its
gradient.

In Section~\ref{sec:otherfuncs}, we go on to consider several examples
of other functionals for physical calculations, including the local
spin-density approximation (LSDA), self-interaction correction (SIC),
density-functional variational perturbation theory, and band-structure
calculations.  We derive the requisite expressions for the
corresponding functionals and their derivatives in the space of a few
pages and thereby show the power and compactness of our formulation
for the treatment of a wide range of single-particle quantum
mechanical problems.

As mentioned in the introduction, our matrix-based formalism allows
the relevant formulae to be literally typed into the computer.
Because these formulae are self-contained, we can make, as illustrated
in the upper-right portion of the figure, a clear distinction between
the expression of the physics itself and the algorithms which search
for the stationary point of the energy functional to achieve
self-consistency.  For concreteness, in Section~\ref{sec:minalgs} we
provide full specification for both a preconditioned
conjugate-gradient minimization algorithm and a new algorithm for
accelerating convergence when working with metallic systems.

Due to our matrix-based formulation, the expressions for the objective
function and its derivatives are built from linear-algebraic
operations involving matrices.  As the lower portion of
Figure~\ref{fig:overview} illustrates, the DFT++ formalism clearly
isolates the software which contains the actual computational kernels.
These kernels therefore may be optimized and parallelized
independently from all other considerations.

Section~\ref{sec:imploptpara} describes these computational
considerations in detail.  In Section~\ref{sec:cppimplem} we discuss
the use of object-oriented languages for linking the underlying
computational kernels with higher level physical expressions.
Section~\ref{sec:FLOPcount} discusses the scaling with physical system
size of the burden for the most time consuming computational kernels.
There are in fact two distinct types of computational kernels, both of
which appear in the lower portion of the figure.

The first type are kernels which implement those few operators in our
formalism that depend on the choice of basis set ($L$, $\calO$, $\I$,
$\J$, $\Idag$, $\Jdag$, defined in Section~\ref{sec:basisdepops}).
These kernels represent the only entrance of basis-set details into
the overall framework.  (Appendix~\ref{appendix:implementpw} provides
the requisite details for plane-wave calculations.)  This allows for
coding of new physics and algorithms without reference to the basis
and for a single higher-level code to be used with ``plug-ins'' for a
variety of different basis sets.  The application of the
basis-dependent operators can be optimized as discussed in Section
\ref{sec:optim} by calling standard packages such as FFTW \cite{fftw}.
Parallelization for the basis-dependent operators is trivial because
they act in parallel on all of the electronic wave functions at once.
Section~\ref{sec:parallel} discusses such parallelization for SMP and
DMP architectures.

Finally, the second class of kernels are basic linear-algebraic
operations (e.g.  matrix multiplication '*', addition '+', subtraction
'-', and Hermitian-conjugated multiplication '\^{}') which do not in
any way depend on the basis set used for the calculation.  As such,
the work of optimization and parallelization for these kernels need
only be performed once.  Section~\ref{sec:optim} presents the two
strategies we employ for these optimizations: blocking of matrix
multiplication and calling optimized linear-algebra packages such as
BLAS3.  Parallelization of these operations is not trivial because
data-sharing or communication is required between processors.  We
detail high performance strategies for dealing with this issue in
Sections~\ref{sec:parallel} for both SMP and DMP architectures.

\section{Lagrangian formalism}
\label{sec:lagr}

The traditional equations of density-functional theory are the
Kohn-Sham equations \cite{KohnSham} for a set of effective
single-particle electronic states $\psirset$.  Below, when we refer to
``electrons'', we are in fact always referring to these effective
electronic degrees of freedom.  The electrostatic or Hartree field
$\phi(r)$ caused by the electrons is traditionally found from solving
the Poisson equation with the electron density derived from these wave
functions as the source term.  The ground-state energy of the system
is then found by minimizing the traditional energy functional, which
ensures the self-consistent solution of the Kohn-Sham equations.  A
great advantage of this variational principle is that first-order
errors in the wave functions lead to only second order errors in the
energy.  However, although not frequently emphasized, errors in
solving the Poisson equation due to the incompleteness of the basis
set used in a calculation may produce a non-variational
(i.e. first-order) error in the energy.

We now consider a new variational principle which ultimately leads to
identical results for complete basis sets, but which places $\psiset$
and $\phi$ on an equal footing and has several advantages in practice.
The central quantity in this principle is the Lagrangian $\LLDA$
introduced in \cite{LAE}, which within the local-density approximation
(LDA), is
\begin{eqnarray}
\LLDA \left(\psirset,\phi(r)\right)
& = & -{1\over 2}\, \sum_i f_i\int d^3r\ 
\psi_i^*(r)\nabla^2\psi_i(r)\nonumber\\
& & + \int d^3r\ V_{ion}(r)\,n(r) +
\int d^3r\ \exc(n(r))\,n(r)\nonumber\\
& & - \int d^3r\ \phi(r)\,\left( n(r) - n_0 \right)
 - {1\over 8\pi}\,\int d^3r\ \|\vec{\nabla}\phi(r)\|^2\,,
\label{eq:lagr}
\end{eqnarray}
where the electron density $n(r)$ is defined in terms of the
electronic states and the Fermi-Dirac fillings $f_i$ as
\begin{equation}
n(r) = \sum_i f_i\, \|\psi_i(r)\|^2\ .
\label{eq:ndef}
\end{equation}
Here and throughout this article we work in atomic units and therefore
have set $\hbar = m_e = e = 1$, where $m_e$ is the electron mass and
$e$ is the charge of the proton.  Finally, the Kohn-Sham electronic
states $\psiset$ must satisfy the orthonormality constraints
\begin{equation}
\int d^3r\ \psi_i^*(r)\,\psi_j(r) = \delta_{ij}\ .
\label{eq:orthonorm}
\end{equation}

Above, $\exc(n)$ is the exchange-correlation energy per electron of a
uniform electron gas with electron density $n$, and $V_{ion}(r)$ is the
potential each electron feels due to the ions.  The constant $n_0$ is
used in calculations in periodic systems as a uniform positive
background that neutralizes the electronic charge density.  The effect
of this background on the total energy is properly compensated when
the Ewald summation is used to compute the interionic interactions.

The following equations, subject to the constraints of
Eq.~(\ref{eq:orthonorm}), specify the stationary point of $\LLDA$,
\begin{eqnarray}
{1\over f_i}{\delta \LLDA  \over \delta \psi_i^*(r)} & = &
0 = \left[-{1 \over 2}\nabla^2 + V_{ion}(r) - \phi(r) +
V_{xc}(r) \right]\psi_i(r) - \epsilon_i\psi_i(r)\ ,
\label{eq:KohnSham}\\
{\delta \LLDA  \over \delta \phi(r)} & = & 0 = -(n(r)-n_0) +
{1\over 4\pi}\nabla^2\phi(r)\,.
\label{eq:Poisson}
\end{eqnarray}
These are seen to be the standard Kohn-Sham eigenvalue equations for
$\psi_i(r)$ and the Poisson equation for the Hartree potential
$\phi(r)$ where the negative sign in the second equation properly
accounts for the negative charge of the electrons.

The behavior of $\LLDA$ is in fact quite similar to that of the
traditional LDA energy functional,
\begin{eqnarray}
\ELDA\left(\psirset\right) & = & -{1\over 2}\, \sum_i f_i\int d^3r\
\psi_i^*(r)\nabla^2\psi_i(r) + \int d^3r\ V_{ion}(r)\,n(r)\nonumber\\
& & +\ \int d^3r\ \exc(n(r))\,n(r)
+ {1\over 2}\int d^3r \int d^3r'\ {n(r)n(r'\,) \over \|r-r'\|}\,.
\label{eq:ELDA}
\end{eqnarray}
First, as shown in \cite{wavelets}, evaluation of
$\LLDA\left(\psirset,\tilde\phi(r)\right)$, where $\tilde\phi$ is the
solution of the Poisson equation, recovers the value of the
traditional energy functional.  Moreover, the derivatives of $\LLDA$
and $\ELDA$ are also equal at $\tilde\phi$.  This result follows by
considering a variation of the equality $\ELDA\left(\psirset\right) =
\LLDA\left(\psirset,\tilde\phi(r)\right)$, which expands into
\begin{eqnarray*}
\sum_i \int d^3r\left(
{\delta \ELDA \over \delta\psi_i(r)}\,\delta\psi_i(r) + 
{\delta \ELDA \over \delta\psi_i^*(r)}\,\delta\psi_i^*(r) \right) & = &
\sum_i \int d^3r\left(
{\delta \LLDA \over \delta\psi_i(r)}\,\delta\psi_i(r) +
{\delta \LLDA \over \delta\psi_i^*(r)}\,\delta\psi_i^*(r)\right)\\
& & +\ \int d^3r\,{\delta \LLDA \over \delta\phi(r)}\,\delta\tilde\phi(r)\,.
\end{eqnarray*}
Because Poisson's equation (Eq.~(\ref{eq:Poisson})) is the condition
that the functional derivative of $\LLDA$ with respect to $\phi$
vanishes, the last term on the right-hand side is zero when
$\phi=\tilde\phi$.  Therefore, the functional derivatives of $\ELDA$
and $\LLDA$ with respect to the electronic states $\psiset$ are equal
when evaluated at $\tilde\phi(r)$.  Finally, the critical points of
$\LLDA$ are in one-to-one correspondence with the minima of $\ELDA$.
The reason is that (i) for fixed $\psiset$, there is a unique
$\tilde\phi$ (up to a choice of arbitrary constant) which solves the
Poisson equation because $\LLDA$ as a function of $\phi$ is a
negative-definite quadratic form, and (ii) as we have just seen, at
such points the derivatives with respect to $\psiset$ of $\ELDA$ and
$\LLDA$ are identical.

One advantage of placing $\phi$ and $\psiset$ on an equal footing is
that now errors in the Lagrangian are second-order in the errors of
both the wave functions {\em and} the Hartree field.  Additionally, as
a practical matter, one has greater flexibility in locating the
stationary point.  Rather than solving for the optimal $\phi$ at each
value of $\psiset$, as is done in traditional DFT methods, one has the
option of exploring in both $\psiset$ and $\phi$ simultaneously.
However, some care in doing this is needed, because the stationary
points of $\LLDA$ are not extrema but saddle points.  (Note that the
first term of Eq.~(\ref{eq:lagr}), the kinetic energy, is unbounded
above, whereas the last term, the Hartree self-energy, is unbounded
below.)  This saddle has a particularly simple structure, and a method
to exploit this is outlined in \cite{torkelarias}.

Finally, by allowing $\phi$ to be a free variable, we have rendered
local all interactions among the fields.  One great formal advantage
is that the subtle mathematical issues in periodic systems arising
from the long-range nature of the Coulomb interaction no longer
require special treatment.  For example, the choice of the
neutralizing background $n_0$ in periodic systems is straightforward
and is treated in detail in Section~\ref{sec:eHinter}.  Because of
this and the aforementioned advantages, we will work in the Lagrangian
framework throughout this article.

\section{Basis-set independent matrix formulation}
\label{sec:matrixnotation}

Our basis-set independent matrix formalism allows us to express the
structure of any single-particle quantum theory in a compact and
explicit way.  In this section, we apply it to the Lagrangian of
Eq.~(\ref{eq:lagr}) which contains energetic terms and non-linear
couplings that are common to all such theories.

To make progress, we first must deal with the fact that the Lagrangian
is a function of continuous fields.  When we perform a computation, we
are forced to represent these fields in terms of expansions within a
finite basis set.  Denoting our basis functions as $\{b_\alpha(r)\}$,
where Greek letters index basis functions, we expand the wave
functions and Hartree potential in terms of expansion coefficients
$C_{\alpha i}$ and $\phi_\alpha$ through
\begin{equation}
\psi_i(r) = \sum_\alpha b_\alpha(r)\, C_{\alpha i}\ \ , \ \ \phi(r) =
\sum_\alpha b_\alpha(r)\, \phi_\alpha\ .
\label{eq:defCphi}
\end{equation}
Typical and popular choices of basis functions are plane waves
(i.e. Fourier modes) \cite{RMP}, finite-element functions
\cite{finiteel}, multiresolution analyses \cite{wavelets}, or Gaussian
orbitals \cite{gauss}.  Once a basis set has been chosen,
$\LLDA$ becomes an explicit function of the finite set of variables
$C_{\alpha i}$ and $\phi_\alpha$.

In addition to the basis set itself, we require a grid of points $p$
in real space covering the simulation cell.  This grid is necessary
for a number of operations, such as for computing the values of the wave
functions or the electron density in real-space and for computing the
exchange-correlation energy density $\exc(n(r))$ of
Eq.~(\ref{eq:lagr}), which is a non-algebraic function of the electron
density $n(r)$ and can only be computed point by point on the
real-space grid.

Our aim is to find a compact, matrix-based notation that works in the
space of expansion coefficients $C_{\alpha i}$ and $\phi_\alpha$ and
is thus applicable to any calculation within any basis set.  In the
course of doing so, we will be able to clearly identify which parts of
our formalism require information about the particular basis that is
chosen and which parts are completely general and independent of this
choice.  In addition, when we have arrived at a matrix-based notation,
it will be clear that only a few fundamental types of computational
kernels are needed to perform the calculation, so that parallelization
and optimization need only address themselves to these few kernels.
This provides a great boon for portability, ease of programming, and
extensibility to new physical scenarios.

In the discussion below, we describe our formalism only for the case
of local ionic potentials.  The use of non-local potentials (e.g. for
the important case of pseudopotential calculations) results in only
minor changes that are addressed in Appendix~\ref{appendix:nlpots}.
Furthermore, for periodic systems, we will be working with only a
single $k$-point at $k=0$, as is evident from the choice of expansion
in Eq.~(\ref{eq:defCphi}).  We work at $k=0$ in order to keep the
mathematical expressions as transparent as possible.  The minor
extensions required to accommodate non-zero and multiple $k$-points
are straightforward and are dealt with in
Appendix~\ref{appendix:kpts}.

\subsection{Basis-dependent operators}
\label{sec:basisdepops}

In this section we describe all the operators in our formalism that
depend on the basis set chosen for the calculation.  We
will see that there are a small number of such operations, and that we
can easily separate their role from the rest of the formalism.

The first two operators involve matrix elements of the identity and
the Laplacian between pairs of basis functions.  Specifically, we
define
\begin{eqnarray}
\calO_{\alpha,\beta} & \equiv & \int d^3r\ 
b^*_\alpha(r)\,b_\beta(r)\label{eq:Odef}\,,\\
L_{\alpha,\beta} & \equiv & \int d^3r\ 
b^*_\alpha(r)\,\nabla^2\,b_\beta(r)\label{eq:Ldef}\,.
\end{eqnarray}
We call these operators the overlap and Laplacian respectively.  Note
that for orthonormal bases, we have $\calO=I$ where $I$ is the
identity matrix.

The integrals of the basis functions are the components of the column
vector $s$,
\begin{equation}
s_\alpha \equiv \int d^3r\ b^*_\alpha(r)\,.
\label{eq:sdef}
\end{equation}
For periodic systems, we use the vector $s$ to define a new operator
$\Obar$ through
\begin{equation}
\Obar \equiv \calO - {ss^\dag \over \Omega}\,,
\label{eq:Obardef}
\end{equation}
where $\Omega$ is the volume of the periodic supercell.  For
calculations in systems without boundaries, the volume $\Omega$ is
infinite so that $\calO=\Obar$.  The chief use of $\Obar$ is for
solving the Poisson equation in periodic systems where divergences due
to the long-range Coulomb interaction must be avoided.  The automatic
avoidance of such divergences and the proper choice of $n_0$ in
Eq.~(\ref{eq:lagr}) both follow directly from the nature of the
Lagrangian as will be discussed in Section~\ref{sec:eHinter}.

The next four operators involve the values of the basis functions on
the points $p$ of the real-space grid introduced above.  The {\em
forward transform} operator $\I$ allows for changing
representation from the space of expansion coefficients to the space
of function values on the real-space grid.  Specifically, given a
basis function $\alpha$ and a grid point $p$, we define
\begin{equation}
\I_{p \alpha} = b_{\alpha}(p)\,.
\label{eq:Idef}
\end{equation}
Thus the $\alpha$th column of $\I$ consists of the values of the
$\alpha$th basis function on the points of the real-space grid.

Next, it is at times necessary to find the expansion coefficients for
a function given its values on the real-space grid.  We denote this
linear {\em inverse transform} by $\J$.  In implementations where the
number of grid points $p$ is equal to the number of basis functions,
the natural choice is to take $\J = \I^{-1}$ (e.g., plane-wave
basis sets).  However, this is not necessary: in some applications,
one may choose to use a very dense real-space grid which has more
points than the number of basis functions.  Hence, we keep the formal
distinction between $\J$ and $\I^{-1}$.  We will also require two {\em
conjugate} transforms, which are the Hermitian conjugates $\Idag$ and
$\Jdag$.

The final mathematical object that depends on the basis set involves
the ionic potential $V_{ion}(r)$.  We define a column vector $V_{ion}$
whose components are the integrals
\begin{equation}
(V_{ion})_\alpha \equiv \int d^3r\ b^*_\alpha(r)\,V_{ion}(r)\,,
\label{eq:Viondef}
\end{equation}
which encodes overlaps of the ionic potential with the basis
functions.  We will use $V_{ion}$ when evaluating the electron-ion
interaction energy in Section~\ref{sec:eiinter}.

\subsection{Identities satisfied by the basis-dependent operators}
\label{sec:OLIJidents}

Although the operators $\calO$, $\Obar$, $L$, $\I$, $\J$, $\Idag$, and
$\Jdag$ depend on the choice of basis, they satisfy various identities
which will prove important below.  In addition to their formal
properties, these identities allow for verification of the
implementation of these operators.

The most important identity involves the constant function.  To
represent the constant function on the grid, we define the column
vector {\bf 1} as having the value of unity on each grid point $p$:
{\bf 1}$_p$ = 1.  Many basis sets can represent this function exactly
(e.g. plane waves or finite-element sets).  For such bases, for all
points $r$ in the simulation cell, we must have the identity
\begin{equation}
\sum_\alpha (\J{\bf 1})_{\alpha}\,b_\alpha(r) = 1\,.
\label{eq:J1ident}
\end{equation}
Evaluating this identity on the real-space grid yields
\begin{equation}
\I\J{\bf 1} = {\bf 1}\,.
\label{eq:IJident}
\end{equation}
For basis sets that can not represent the constant function exactly,
the identity of Eq.~(\ref{eq:IJident}) and the ones below should hold
approximately in the regions described by the basis.

According to Eq.~(\ref{eq:J1ident}), the vector $\J{\bf 1}$ specifies
the coefficients of the expansion of the constant function.  Using the
integrals $s$ of Eq.~(\ref{eq:sdef}), we can see that
\begin{eqnarray*}
s_\alpha & = & \int d^3r\ b^*_\alpha(r)\\
& = & \int d^3r\ b^*_\alpha(r)\left(\sum_\beta(\J{\bf
1})_{\beta}b_\beta(r)\right)\\
& = & \left(\calO\J{\bf 1}\right)_\alpha\,.
\end{eqnarray*}
Thus we have that $s=\calO\J{\bf 1}$.  We can also derive the
normalization condition
\begin{equation}
s^\dag\J{\bf 1}=\int d^3r\ \sum_\alpha b_\alpha(r)(\J{\bf
1})_\alpha = \int d^3r\ 1 = \Omega\,,
\label{eq:sident}
\end{equation}
where $\Omega$ is the volume in which the calculation is performed.

When solving Poisson's equation for the electrostatic potential
(Eq.~(\ref{eq:Poisson})), we must take special care regarding the null
space of the Laplacian operator $L$, which is the space of constant
functions.  Integrating the identity $\nabla^2 1=0$ against the
complex conjugate of each basis function yields
\begin{equation}
L\J{\bf 1} = 0\,.
\label{eq:LJ1ident}
\end{equation}
We use this identity when dealing with the Poisson equation in
periodic systems to avoid divergences due to the long-range
nature of the Coulomb interaction.

\subsection{Basis-independent expression for the Lagrangian}
\label{sec:basisindepexprlagr}

We now use the above operators to express the Lagrangian of
Eq.~(\ref{eq:lagr}) in a matrix-based, basis-independent manner.  We
begin by introducing two operators, ``diag'' and ``Diag''. The
operator diag converts a square matrix $M$ into a column vector
containing the diagonal elements of the matrix.  The operator Diag
converts a vector $v$ into a diagonal matrix with the components of
$v$ on its diagonal.  In terms of components, we have that
\begin{equation}
(\mbox{diag}\ M)_\alpha = M_{\alpha\alpha}\ \ , \ \ \ (\mbox{Diag}\ 
v)_{\alpha\beta} = v_\alpha\delta_{\alpha\beta}\,,
\label{eq:diagdef}
\end{equation}
where $\delta$ is the Kronecker delta.  Thus, diag Diag $v$ = $v$ for
any vector $v$ whereas Diag diag $M$ = $M$ if and only if $M$ is a
diagonal matrix.  Two useful identities involving these operators are
\begin{equation}
(\mbox{diag}\ M)^\dag\, v = \mbox{Tr}(M^\dag\,\mbox{Diag}\ v)\ \ , \ \ \ 
v^\dag\,(\mbox{diag}\ M) = \mbox{Tr}(\,(\mbox{Diag}\ v)^\dag\, M)\,,
\label{eq:diagidents}
\end{equation}
where $^\dag$ indicates Hermitian or complex-conjugated transposition.

Next, if we regard the expansion coefficients $C_{\alpha i}$ as a
matrix whose $i$th column contains the expansion coefficients of the
$i$th wave function (Eq.~(\ref{eq:defCphi})), and we also define the
diagonal matrix of Fermi fillings $F_{ij}=f_i\delta_{ij}$, it is easy
to see that
\begin{equation}
P = CFC^\dag
\label{eq:Pdef}
\end{equation}
is the representation of the single-particle density matrix in the
space of basis functions.

Before considering the Lagrangian itself, we will also need
expressions for the electron density $n(r)$ which appears in most of
the terms of the Lagrangian of Eq.~(\ref{eq:lagr}).  We define a
vector $n$ whose components are the values of the electron density on
the points $p$ of the real-space grid.  Specifically,
\begin{eqnarray}
n_p \equiv n(p) & = & \sum_i f_i \|\psi_i(p)\|^2  = 
\sum_i f_i \|\left(\I C\right)_{pi}\|^2\nonumber\\
& = & \sum_i \left(\I C\right)^*_{pi}f_i\left(\I C\right)_{pi} =
\left((\I C)F(\I C)^\dag\right)_{pp}\,,\nonumber
\end{eqnarray}
whence we arrive at the identity defining the vector $n$
\begin{equation}
n = \mbox{diag}\left(\I P\Idag\right)\,.
\label{eq:nexpr}
\end{equation}
Given the values of the electron density on the real-space grid, we
use the inverse transform $\J$ to find the expansion coefficients of
$n(r)$ in terms of the basis functions.  This vector of coefficients
is just $\J n$.

Armed with these few tools, we now proceed to write the various
energetic terms of the Lagrangian in the matrix language developed
above.

\subsubsection{Kinetic energy}

The kinetic energy $T$ can be transformed into the matrix language by
using the expansion coefficients $C$ of Eq.~(\ref{eq:defCphi}) and by
using the definition of the Laplacian $L$ of
Eq.~(\ref{eq:Ldef}):
\begin{eqnarray}
T & \equiv & -{1\over 2}\sum_i f_i \int d^3r\
\psi_i^*(r)\nabla^2\psi_i(r) =  -{1\over 2}\sum_i f_i
\sum_{\alpha,\beta} C^*_{\alpha i}L_{\alpha\beta}C_{\beta i}\nonumber\\
& = & -{1\over 2}\mbox{Tr}\left(FC^\dag LC\right) = -{1\over
2}\mbox{Tr}\left(LP\right)\,,
\label{eq:Texpr}
\end{eqnarray}
where the last two equivalent expressions are related by the cyclic
property of the trace.  Thus, we are able to write the kinetic energy
explicitly as a function of the density matrix $P$ of
Eq.~(\ref{eq:Pdef}).

\subsubsection{Electron-ion interaction}
\label{sec:eiinter}

Since the electron density $n(r)$ is real, we may write the
electron-ion interaction as
\begin{eqnarray}
E_{e-i} & \equiv & \int d^3r\ n^*(r)\,V_{ion}(r) = \sum_\alpha \int
d^3r (\J n)_\alpha^* b^*_\alpha(r)\,V_{ion}(r)\nonumber\\
& = & (\J n)^\dag V_{ion} = \mbox{Tr}\left( \Idag\left[\mbox{Diag }\Jdag
V_{ion}\right]\I P\right),
\label{eq:Eeiexpr}
\end{eqnarray}
where we have used the definition of $V_{ion}$ from
Eq.~(\ref{eq:Viondef}) and have used Eqs.~(\ref{eq:nexpr}) and
(\ref{eq:diagidents}) to rewrite this interaction in terms of $P$.

\subsubsection{Exchange-correlation energy}

Given the vector $n$ of electron-density values on the grid, we can
evaluate the exchange-correlation energy per particle at each grid
point $p$ through $\exc(n(p))$.  We collect these values into a vector
$\exc(n)$.  We then inverse transform this vector and the electron
density vector, and we use the overlap operator to arrive at
\begin{eqnarray}
E_{xc} & \equiv & \int d^3r\ n^*(r)\,\exc(n(r))\nonumber\\ & =
& \left(\J n\right)^\dag\calO\left(\J\exc(n)\right) =
\mbox{Tr}\left(\Idag\left[\mbox{Diag }\Jdag\calO\J\exc(n)\right]\I P\right)\,,
\label{eq:Excexpr}
\end{eqnarray}
where we again have conjugated the electron density for ease of formal
manipulations.  The derivation of the final expression in terms of $P$
uses Eq.~(\ref{eq:nexpr}).

\subsubsection{Hartree self-energy}
\label{sec:Hselfener}

The self-energy of the Hartree field can be written as
\begin{equation}
E_{H-H} \equiv -{1\over 8\pi} \int d^3r\ \|\vec\nabla\phi(r)\|^2 =
{1\over 8\pi} \int d^3r\ \phi^*(r)\nabla^2\phi(r) = {1\over
8\pi}\phi^\dag L\phi\,,
\label{eq:EHHexpr}
\end{equation}
where we have first integrated by parts and then substituted the
expansion coefficients $\phi$ of Eq.~(\ref{eq:defCphi}).  The complex
conjugation of the real-valued function $\phi(r)$ allows for the
simplicity of the final expression.

\subsubsection{Electron-Hartree interaction}
\label{sec:eHinter}

The interaction of the electron density $n(r)$ and Hartree potential
$\phi(r)$ can be written as
\begin{equation}
E_{e-H} \equiv - \int d^3r\ \left(n(r)-n_0\right)^*\phi(r) = -
\left[\J (n-n_0{\bf 1})\right]^\dag\calO\phi\,.
\label{eq:EeHdef}
\end{equation}
The proper choice of $n_0$ for periodic systems can be found by
noting that the Hartree self-energy $E_{H-H}$ of
Eq.~(\ref{eq:EHHexpr}) has no dependence on the projection of $\phi$
onto the null space of $L$ which, as we saw in
Section~\ref{sec:OLIJidents}, lies along the vector $\J{\bf 1}$.
Thus, for the Lagrangian of Eq.~(\ref{eq:lagr}) to have a
saddle-point, there can be no coupling of $n(r)-n_0$ with the
projection of $\phi$ along $\J{\bf 1}$.  That is, we must have
$\left[\J (n-n_0{\bf 1})\right]^\dag\calO \cdot\J{\bf 1} = 0$.  The
identities of Section~\ref{sec:OLIJidents} then lead to the choice
$n_0=(\J n)^\dag s/\Omega$.  Our final expression for $E_{e-H}$ is
thus given by
\begin{equation}
E_{e-H} = - (\J n)^\dag\left(\calO-{ss^\dag \over\Omega}\right)\phi =
- (\J n)^\dag\Obar\phi = -\mbox{Tr}\left(\Idag\left[\mbox{Diag }
\Jdag\Obar\phi\right]\I P\right)\,.
\label{eq:EeHexpr}
\end{equation}

\subsubsection{Complete Lagrangian}

Summing all the contributions above, we arrive at two equivalent
expressions for the Lagrangian $\LLDA$,
\begin{eqnarray}
\LLDA & = & -{1\over 2}\mbox{Tr}\left(FC^\dag
LC\right) + \left(\J n\right)^\dag \left[V_{ion} + 
\calO\J\exc(n) - \Obar\phi\right]
+ {1\over 8\pi}\phi^\dag L\phi
\label{eq:lagrexpr1}\\
& = & -{1\over 2}\mbox{Tr}\left(LP\right) + {1\over 8\pi}\phi^\dag
L\phi \nonumber\\
& & +\ \mbox{Tr}\left(\Idag\mbox{Diag}\left[
\Jdag V_{ion} + \Jdag\calO\J\exc(n)
- \Jdag\Obar\phi
\right]\I P\right)\,.
\label{eq:lagrexpr2}
\end{eqnarray}
The first, compact form is computationally efficient for evaluating
the Lagrangian as a function of $C$ and $\phi$.  The second form,
written as a function of the density matrix $P$, finds its best use in
the formal manipulations required to find the gradient of the
Lagrangian.

\subsection{Orthonormality constraints}
\label{sec:orthonorm}

The orthonormality constraints of Eq.~(\ref{eq:orthonorm}) are
equivalent to the matrix equation
\begin{equation}
C^\dag\calO C = I\,.
\label{eq:COC}
\end{equation}
If we wish to compute gradients of the Lagrangian with respect to $C$
in order to arrive at the Kohn-Sham equations, we must do so while
always obeying these constraints.

The analytically-continued functional approach \cite{ACprl} deals with
these constraints by introducing a set of expansion coefficients $Y$
which are {\em unconstrained} and which can have any overlap $U$,
\begin{equation}
U = Y^\dag \calO Y.
\label{eq:Udef}
\end{equation}
We also allow for the possibility of subspace rotation, which is a
unitary transformation mapping the subspace of occupied states
$\psiset$ onto itself.  Such a transformation is affected by a
unitary matrix $V$, and we parameterize $V$ as the exponential of a
Hermitian matrix $B$ through
\begin{equation}
V \equiv e^{iB}\ \ \ \mbox{where}\ \ \ B^\dag = B\,.
\label{eq:Bdef}
\end{equation}

The coefficients $C$ are defined as dependent variables through the
mapping
\begin{equation}
C = YU^{-1/2}V^\dag
\label{eq:CYUmhalfVdag},
\end{equation}
which ensures that Eq.~(\ref{eq:COC}) is automatically obeyed, as is
easy to verify by direct substitution.  The density matrix $P$ takes
the following form in terms of $Y$ and $V$,
\begin{equation}
P = CFC^\dag = YU^{-1/2}V^\dag FVU^{-1/2}Y^\dag\,.
\label{eq:PofY}
\end{equation}
In most cases, we simply set $V=I$.  In fact, for the case of constant
fillings, $F=fI$, the unitary matrix $V$ drops out of $P$ completely.
The subspace rotations find their primary use in the study of metallic
or high-temperature systems where the Fermi-Dirac fillings are not
constant, and the rotations allow for greatly improved convergence
rates when searching for the saddle point of the Lagrangian.  This
point is explained in more detail in Section~\ref{sec:minalgs}.

\subsection{Derivatives of the Lagrangian}

Since the most effective modern methods that search for stationary
points require knowledge of the derivative of the objective function,
we will now find the derivative of the Lagrangian of
Eq.~(\ref{eq:lagrexpr1}) or (\ref{eq:lagrexpr2}) with respect to the
variables $Y$ and $\phi$ (and $B$ if subspace rotations are used).
Differentiation with respect to $Y$ is far more complex due to the
orthonormality constraints, and we begin with this immediately.

\subsubsection{Derivative with respect to the electronic states}
\label{sec:Yderiv}

Computing the derivative of the Lagrangian with respect to $Y$ is
intricate, and we break the problem into smaller pieces by first
finding the derivative with respect to the density matrix $P$.  Once
the derivative with respect to $P$ is found, we can use the relation
between $P$ and $Y$ (Eq.~(\ref{eq:PofY})) to find the derivative with
respect to $Y$.

We begin by noting that except for the exchange-correlation energy,
the entire expression of Eq.~(\ref{eq:lagrexpr2}) is linear in the
density matrix $P$.  The exchange-correlation energy is a function of
the electron density $n$, which, through Eq.~(\ref{eq:nexpr}), is also
a function of $P$.  Thus if we consider a differential change $dP$ of
the density matrix, the only term in $d\LLDA$ that needs to be
considered carefully is
\begin{eqnarray*}
n^\dag\Jdag\calO\J d[\exc(n)] & = &
n^\dag\Jdag\calO\J\left[\mbox{Diag }\excprime(n)\right]dn\\
& = & \mbox{Tr}\left\{\Idag\mbox{Diag}\left(
\left[\mbox{Diag }\excprime(n)\right]\Jdag\calO\J n\right)
\I\,dP\right\}.
\end{eqnarray*}
In the above derivation, we have used Eq.~(\ref{eq:nexpr}) to relate
$dn$ to $dP$ as well as the identities of Eq.~(\ref{eq:diagidents}).
The vector $\excprime(n)$ is given by its values on the real-space
grid points $p$ via $(\excprime(n))_p \equiv \excprime(n(p))$.

We can now write the differential of the Lagrangian of
Eq.~(\ref{eq:lagrexpr2}) with respect to $P$ as
\begin{eqnarray}
d\LLDA  & = & \mbox{Tr}\left\{-{1\over 2}L\,dP + 
\Idag\mbox{Diag}\left[ \Jdag V_{ion} + \Jdag\calO\J
\exc(n)\right.\right.\nonumber\\ & & \left.\left. \ \ \ \
\ \ \ \ + \ \left[\mbox{Diag }\excprime(n)\right]\Jdag\calO\J n
- \Jdag\Obar\phi\right]\I\,dP\right\}\nonumber\\ &
\equiv & \mbox{Tr}\left(H\,dP\right)\,,
\label{eq:HdP}
\end{eqnarray}
where the single-particle Kohn-Sham Hamiltonian operator $H$ is
given by
\begin{eqnarray}
H & = & -{1\over 2}L + \Idag [\mbox{Diag }V_{sp}]\,\I\,, \ \
\mbox{where}\nonumber\\
V_{sp} & = & \Jdag V_{ion} + \Jdag\calO\J \exc(n) +
[\mbox{Diag }\excprime(n)]\Jdag\calO\J n - \Jdag\Obar\phi\,.
\label{eq:Hspdef}
\end{eqnarray}
The single-particle Hamiltonian is the sum of a kinetic operator and a
local single-particle potential $V_{sp}$ (a vector of numbers on the
real-space grid specifying the values of the potential).

Eq.~(\ref{eq:HdP}) has conveniently separated out the physical
description of the system, the Hamiltonian $H$, from the variation
$dP$ which we now compute.  Differentiating the relation
$U^{-1/2}U^{1/2}=I$, we find that
\[
d[U^{-1/2}] = - U^{-1/2}d[U^{1/2}]U^{-1/2},
\]
and we use this to express the variation of the density matrix of
Eq.~(\ref{eq:PofY}) as
\begin{eqnarray*}
dP & = & (dY)U^{-1/2}V^\dag FVU^{-1/2}Y^\dag +
 YU^{-1/2}V^\dag FVU^{-1/2}(dY^\dag)\nonumber\\
& & - \ YU^{-1/2}\left(d[U^{1/2}]U^{-1/2}V^\dag FV +
V^\dag FVU^{-1/2}d[U^{1/2}]\right)U^{-1/2}Y^\dag.
\end{eqnarray*}
We now substitute this expression for $dP$ into Eq.~(\ref{eq:HdP}).
We use the definition of the operator $Q$ (Eq.~(\ref{eq:Qdef}) of
Appendix~\ref{appendix:UmhalfQ}), its relation to $d[U^{1/2}]$
(Eq.~(\ref{eq:UhalfQ})), and the identities which $Q$ satisfies
(Eqs.~(\ref{eq:Qidents})).  After some manipulations involving the
cyclicity of the trace, we arrive at
\begin{eqnarray}
d\LLDA & = & \mbox{Tr}\left[dY^\dag \left({\partial
\LLDA \over \partial Y^\dag}\right) + \left({\partial
\LLDA \over \partial Y^\dag}\right)^\dag dY\right],\
\mbox{where}\nonumber\\
\left({\partial \LLDA  \over \partial Y^\dag}\right) & \equiv & 
\left(I-\calO CC^\dag\right)HCFVU^{-1/2} + 
\calO CVQ\left(V^\dag[\tilde{H},F]V\right)\,,\ \mbox{and}\nonumber\\
\tilde{H} & \equiv &  C^\dag HC\,,
\label{eq:dLdYdag}
\end{eqnarray}
where $\tilde{H}$ is the subspace Hamiltonian and contains matrix
elements of the Hamiltonian $H$ among the wave functions $\psirset$.
Square brackets denote the commutator, $[a,b]\equiv ab-ba$.  Physical
interpretation of the terms in Eq.~(\ref{eq:dLdYdag}) is provided in
Section~\ref{sec:KohnShamPoisson}.

Finally, since $Y$ and $Y^\dag$ are not independent, we can simplify
the expression for the differential of $\LLDA$ to
\[
d\LLDA  = \mbox{2 Re Tr}\left[dY^\dag \left({\partial
\LLDA  \over \partial Y^\dag}\right)\right]\,,
\]
where Re denotes the real part of its argument.

\subsubsection{Derivative with respect to the Hartree field}

Since the Lagrangian in Eq.~(\ref{eq:lagrexpr1}) is quadratic in
$\phi$, its derivative with respect to $\phi$ may be readily
calculated.  However, to arrive at a symmetric expression for the
derivative in terms of $\phi$ and $\phi^\dag$, we note that the linear
dependence on $\phi$, given by $z \equiv (\J n)^\dag\Obar\phi$, is a
real number because both $n(r)$ and $\phi(r)$ are real in
Eq.~(\ref{eq:EeHdef}).  For convenience, we rewrite this as
$(z+z^*)/2$, which is an equivalent expression symmetric in $\phi$ and
$\phi^\dag$.  By using this, we compute the variation of $\LLDA$ and
arrive at
\begin{eqnarray}
d\LLDA  & = & d\phi^\dag \left({\partial \LLDA  \over
\partial \phi^\dag }\right) + \left({\partial \LLDA  \over
\partial \phi^\dag }\right)^\dag d\phi\,,\ \mbox{where}\nonumber\\
\left({\partial \LLDA  \over \partial \phi^\dag }\right) &
\equiv & -{1\over 2}\Obar\J n + {1\over 8\pi}L\phi\,.
\label{eq:dLdphi}
\end{eqnarray}
Again, since $\phi$ and $\phi^\dag$ are not independent, we can
express the variation as
\[
d\LLDA  = 2\, \mbox{Re} \left[ d\phi^\dag \left({\partial
\LLDA  \over \partial \phi^\dag }\right)\right]\,.
\]

\subsubsection{Derivative with respect to subspace rotations}

We have parameterized the unitary matrix $V$ of
Eq.~(\ref{eq:CYUmhalfVdag}) by the Hermitian matrix $B$ of
Eq.~(\ref{eq:Bdef}) as $V=e^{iB}$. We will now find the derivative of
$\LLDA$ with respect to $B$.  First, we consider the variation of
$\LLDA$ with respect to those of $V$ and $V^\dag$ by using
Eq.~(\ref{eq:HdP}) as our starting point. Using the definition of $P$
in Eq.~(\ref{eq:PofY}), we have that
\begin{eqnarray*}
d\LLDA  & = & \mbox{Tr}\{H\,dP\}\\ & = &
\mbox{Tr}\left\{HYU^{-1/2}[dV^\dag FV + V^\dag
FdV]U^{-1/2}Y^\dag\right\}\\ & = & \mbox{Tr}\left\{\tilde{H}'[dV^\dag FV +
V^\dag FdV]\right\}\,,
\end{eqnarray*}
where $\tilde{H}'=U^{-1/2}Y^\dag HYU^{-1/2}$.
Differentiating the identity $V^\dag V=I$ leads to $dV^\dag = -V^\dag
dV V^\dag$ which allows us to write
\[
d\LLDA  = \mbox{Tr}\left\{[\tilde{H},F]dVV^\dag\right\}\,,
\]
where again $\tilde{H}=C^\dag HC$ is the subspace Hamiltonian.

We place the eigenvalues of $B$ on the diagonal of a diagonal matrix
$\beta$ and place the eigenvectors of $B$ in the columns of a unitary
matrix $Z$.  Thus $B=Z\beta Z^\dag$ and $Z^\dag Z=ZZ^\dag =I$.  We
now use the result of Eq.~(\ref{eq:dfU}) of
Appendix~\ref{appendix:UmhalfQ} applied to the case $V=f(B)=e^{iB}$ to
arrive at the following result relating $dV$ to $dB$:
\[
(Z^\dag dV Z)_{nm} = (Z^\dag dB Z)_{nm} \cdot \left\{
\begin{array}{cc}
ie^{i\beta_n} & \mbox{if }m = n\\
\left[{e^{i\beta_m}-e^{i\beta_n} \over \beta_m-\beta_n}\right]
& \mbox{if }m \neq n
\end{array}
\right. .
\]
Using this and the fact that $V^\dag = Ze^{-i\beta}Z^\dag$, we have
that
\begin{eqnarray*}
d\LLDA  & = & \mbox{Tr}\left\{[\tilde H,F]Z(Z^\dag dVZ)Z^\dag
V^\dag\right\}\\
& = & \mbox{Tr}\left\{Z^\dag [\tilde H,F]Z(Z^\dag dVZ)e^{-i\beta}\right\}\\
& = & \sum_{n,m} (Z^\dag [\tilde
H,F]Z)_{nm}(Z^\dag dBZ)_{mn}\cdot\left\{
\begin{array}{cc}
i & \mbox{if }m = n\\
\left[{e^{i\beta_m-i\beta_n}-1 \over \beta_m-\beta_n}\right]
& \mbox{if }m \neq n
\end{array}
\right. .
\end{eqnarray*}
We define the operator $R(A)$ acting on a general matrix $A$ via
\[
(Z^\dag R(A)Z)_{nm} \equiv (Z^\dag AZ)_{nm}\cdot\left\{
\begin{array}{cc}
i & \mbox{if }m = n\\
\left[{e^{i\beta_m-i\beta_n}-1 \over \beta_m-\beta_n}\right]
& \mbox{if }m \neq n
\end{array}
\right. .
\]
This allows us to write the variation of $\LLDA$ as
\[
d\LLDA  = \mbox{Tr}\left\{Z^\dag R\left([\tilde
H,F]\right)ZZ^\dag dBZ\right\} = \mbox{Tr}\left\{R\left([\tilde
H,F]\right)dB\right\}\,,
\]
so that the derivative of $\LLDA$ with respect to $B$ is
\begin{equation}
{\partial \LLDA  \over \partial B} = R\left([\tilde H,F]\right)\,.
\label{eq:dLdB}
\end{equation}

\subsection{Kohn-Sham and Poisson equations}
\label{sec:KohnShamPoisson}

The Kohn-Sham and Poisson equations (Eqs.~(\ref{eq:KohnSham}) and
(\ref{eq:Poisson})) are obtained by setting the derivative of the
Lagrangian with respect to $Y$ and $\phi$ to zero.  This results in
the two equations
\begin{eqnarray}
\left({\partial \LLDA  \over \partial Y^\dag}\right) & = & 0 =
\left(I-\calO CC^\dag\right)HCFVU^{-1/2} +
\calO CVQ(V^\dag[\tilde{H},F]V)\,,\label{eq:KSexpr}\\
\left({\partial \LLDA  \over \partial \phi^\dag }\right) &
= & 0 = -{1\over 2}\Obar\J n + {1\over 8\pi}L\phi\,.
\label{eq:Poissonexpr}
\end{eqnarray}
Eq.~(\ref{eq:KSexpr}) states the stationarity of the Lagrangian with
respect to variations of the wave-function coefficients $Y$, and we
examine it first.

We define the projection operator $\rho = \calO CC^\dag$ which
satisfies $\rho^2=\rho$ and which projects onto the subspace of
occupied states $\psiset$ used in the calculation.  Its complement
$\bar\rho = I-\rho$ projects onto the orthogonal subspace spanned by
the unoccupied states.  By multiplying Eq.~(\ref{eq:KSexpr}) on the
left by $\bar\rho$ and assuming that none of the Fermi fillings are
zero, we find that
\[
\bar\rho HC = 0\,.
\]
This reproduces the well known condition that at the stationary
point, the Hamiltonian must map the occupied subspace onto itself.

Conversely, we can project Eq.~(\ref{eq:KSexpr}) onto the occupied
subspace by multiplying on the left by $C^\dag$.  This, combined with
the fact that $Q$ is an invertible linear operator, leads to the
condition
\[
[\tilde H,F] = 0\,.
\]
Given that $F$ is a diagonal matrix, for arbitrary fillings, the
subspace Hamiltonian $\tilde H$ also must be diagonal: the states
$\psiset$ must be eigenstates of $H$ with eigenvalues $\epsilon_i$, as
we have explicitly written in Eq.~(\ref{eq:KohnSham}).  However, if a
pair of states $\psi_i$ and $\psi_j$ have degenerate fillings,
$f_i=f_j$, then $\tilde H_{ij}$ need not be zero.  Converting such
degenerate cases into the conventional diagonal representation
requires a further unitary rotation, which, however, is not required
for stationarity.

The second condition for stationarity, Eq.~(\ref{eq:Poissonexpr}),
rearranges into
\begin{equation}
L\phi = 4\pi\Obar\J n\,.
\label{eq:Poissonexpr2}
\end{equation}
We have arrived at the Poisson equation for the Hartree potential
$\phi$ generated by the electron density $n$ (the negative electronic
charge is reflected by the positive coefficient of the right-hand
side).  Since we have explicitly projected out the null-space of $L$
from the right-hand side, we may invert $L$ and find the solution to
the Poisson equation
\begin{equation}
\phi = 4\pi L^{-1}\Obar\J n\,.
\label{eq:Poissonsolution}
\end{equation}

Finally, if we substitute the result of Eq.~(\ref{eq:Poissonsolution})
for $\phi$ into our Lagrangian, we find the LDA energy functional
(cf. Eq.~(\ref{eq:ELDA})):
\begin{equation}
\ELDA = -{1\over 2}\mbox{Tr}\left(FC^\dag LC\right) +
\left(\J n\right)^\dag\left[V_{ion} + \calO\J\exc(n)
- {1\over 2}\Obar\left(4\pi L^{-1}\Obar\J n\right)\right].
\label{eq:ELDAexpr}
\end{equation}

\subsection{Expressions for Lagrangian and derivatives:  summary}
\label{sec:lagrderivs}

We now collect the expressions for the LDA Lagrangian and its
derivatives in one place.  As we will see in
Section~\ref{sec:imploptpara}, formulae in the DFT++ notation
translate directly into lines of computer code, so that we will also
be specifying the computational implementation of the Lagrangian.  In
addition, given the Lagrangian and its derivatives, we can apply any
suitable algorithm to find the stationary point
(Section~\ref{sec:minalgs}).

The key expressions are
\begin{center}
\framebox[0.95\textwidth]{\parbox{0.95\textwidth}{
\begin{eqnarray*}
\LLDA (Y,\phi,B) & = & -{1\over 2}\mbox{Tr}\left(FC^\dag LC\right)
+\left(\J n\right)^\dag\left[V_{ion} + \calO\J\exc(n) -
\Obar\phi\right] +{1\over 8\pi}\phi^\dag L\phi\,,\\
{\partial \LLDA  \over
\partial Y^\dag} & = & \left(I-\calO CC^\dag\right)HCFVU^{-1/2} +
\calO CVQ\left(V^\dag[\tilde{H},F]V\right)\,,\\
{\partial \LLDA  \over \partial
\phi^\dag} & = & -{1\over 2}\Obar\J n + {1\over8\pi}L\phi\ \ ,\ \ 
{\partial \LLDA  \over \partial B}\, = \, R\left([\tilde{H},F]\right)\,,\\
H & = & -{1\over 2}L + \Idag \left[\mbox{Diag
}V_{sp}\right]\I\ \ ,\ \ \tilde{H}\,=\,C^\dag H C\,.
\end{eqnarray*}}}
\end{center}
As discussed in Section~\ref{sec:lagr}, the value of $\LLDA$ and its
$Y$ and $B$ derivatives are equal to the value and respective
derivatives of the energy $\ELDA$ of Eq.~(\ref{eq:ELDAexpr}) when we
evaluate the Lagrangian-based quantities at the solution of the
Poisson equation.  Therefore, the above expressions can also be used
to find the derivatives of $\ELDA$, a fact that we will exploit in
Section~\ref{sec:minalgs}.

\section{DFT++ specification for various \abi techniques}
\label{sec:otherfuncs}

In the previous section, we presented a detailed derivation of the
expression for the LDA Lagrangian and its derivatives in the DFT++
formalism.  We believe that the community of physicists and chemists
using this and other general-functional methods should use this
formalism for the communication of new energy functionals and
comparisons among them.

From a physicist's or chemist's viewpoint, which we adopt in this
section, linear algebra and matrices are the settings in which quantum
mechanical computations must be performed.  Therefore, they are the
fundamental objects in the new formalism.  This is in contrast with
the Dirac notation, which while an excellent conceptual tool for
studying quantum problems, can never be used to carry out an actual
calculation: matrix elements of bras and kets must first be found
before an actual computation can proceed.

Once an expression for an energy functional is found, its derivative
is found by straightforwardly differentiating it with respect to the
matrices of independent variables.  Armed with expressions for the
functional and its derivative, the methods discussed in
Section~\ref{sec:minalgs} can then be applied to achieve
self-consistency.

In this spirit, we now present a few examples of energy functionals.
Some are extensions of the LDA, while others are similar to the LDA
only in that they involve the study of single-particle systems.  In
all cases, our aim will be to show how quickly and easily we can find
the requisite expressions for the appropriate functional and its
derivative.

\subsection{Local spin-density approximation (LSDA)}

The most straightforward generalization of the LDA approximation is to
allow for spin-up and spin-down electrons to have different wave
functions but to still treat exchange-correlation energies in a local
approximation.  Specifically, the exchange-correlation energy per
particle at position $r$ is now a function of both the spin-up and
spin-down electron densities, $n_\uparrow(r)$ and $n_\downarrow(r)$,
and the total LSDA exchange-correlation energy is given by
\[
E_{xc} = \int d^3r\
n(r)\,\exc(n_\uparrow(r),n_\downarrow(r))\ ,
\]
where $n(r)=n_\uparrow(r) + n_\downarrow(r)$ is the total electron
density.  The LSDA has been found to show substantial improvements over
the LDA for atomic and molecular properties
\cite{GunnarsonLundqvist,ParrYang} since the spin of the electrons is
explicitly dealt with.

The changes required in the expressions of the Lagrangian and its
derivatives in order to incorporate the LSDA are straightforward and
easy to implement.  We label spin states by $\sigma$, which can take
the value $\uparrow$ or $\downarrow$.  We have spin-dependent
expansion coefficient matrices $C_\sigma$ for the wave functions (cf.
Eq.~(\ref{eq:defCphi})).  Each spin channel has its own fillings
$F_\sigma$ and density matrix $P_\sigma$,
\[
P_\sigma = C_\sigma F_\sigma C_\sigma^\dag\,.
\]
The electron densities $n_\sigma$ and the total electron density $n$
are given by (cf. Eq.~(\ref{eq:nexpr}))
\[
n_\sigma = \mbox{diag}(\I P_\sigma\Idag)\ \ \ \mbox{and}\
\ \ n = \sum_\sigma n_\sigma\ .
\]

The LSDA Lagrangian is given by
\[
{\cal L}_{LSDA}(C_\uparrow,C_\downarrow,\phi) = -{1\over
2}\sum_\sigma \mbox{Tr}\left(F_\sigma C_\sigma^\dag LC_\sigma\right) +
\left(\J n\right)^\dag
\left[V_{ion} + \calO\J\exc(n_\uparrow,n_\downarrow)
-  \Obar\phi\right] + {1\over 8\pi}\phi^\dag L\phi.
\]
The orthonormal expansion coefficients $C_\sigma$ are found from
unconstrained variables $Y_\sigma$ via
\[
C_\sigma = Y_\sigma U_\sigma^{-1/2} ,\ \mbox{where}\ U_\sigma =
Y_\sigma^\dag \calO Y_\sigma\,,
\]
where, for simplicity, we have set subspace rotations to unity,
$V_\sigma=I$.  Finding the derivatives of the Lagrangian with respect
to the coefficients $Y_\sigma$ follows the analysis of
Section~\ref{sec:Yderiv}.  Each spin channel has a single-particle
Hamiltonian $H_\sigma$ given by
\begin{eqnarray*}
H_\sigma & = & -{1\over 2}L + \Idag [\mbox{Diag
}V_\sigma]\,\I\ ,\ \mbox{where}\\
V_\sigma & = & \Jdag V_{ion} + \Jdag\calO\J\exc(n_\uparrow,n_\downarrow) +
\mbox{Diag}\left[{\partial \exc(n_\uparrow,n_\downarrow) \over
\partial n_\sigma}\right]\Jdag\calO\J n - \Jdag\Obar\phi.
\end{eqnarray*}
The derivative of the Lagrangian with respect to $Y_\sigma$
(cf. Eq.~(\ref{eq:dLdYdag})) is given by
\begin{eqnarray*}
\left({\partial {\cal L}_{LSDA} \over \partial Y_\sigma^\dag}\right) &
= & \left(I-\calO C_\sigma C_\sigma^\dag\right)H_\sigma C_\sigma
F_\sigma U_\sigma ^{-1/2} + \calO C_\sigma
Q([\tilde{H}_\sigma,F_\sigma]),\ \mbox{where}\ \tilde{H}_\sigma \equiv
C_\sigma^\dag H_\sigma C_\sigma.
\end{eqnarray*}

In summary, we have the following expressions for the LSDA Lagrangian
and derivatives
\begin{center}
\framebox[\textwidth]{\parbox{\textwidth}{
\begin{eqnarray*}
{\cal L}_{LSDA}(Y_\uparrow,Y_\downarrow,\phi) & = & -{1\over
2}\sum_\sigma \mbox{Tr}\left(F_\sigma C_\sigma^\dag LC_\sigma\right) +
{1\over 8\pi}\phi^\dag L\phi\\
& & +\left(\J n\right)^\dag
\left[V_{ion} + \calO\J\exc(n_\uparrow,n_\downarrow)
- \Obar\phi\right]\\
{\partial {\cal L}_{LSDA} \over
\partial Y_\sigma^\dag} & = & \left(I-\calO C_\sigma
C_\sigma^\dag\right)H_\sigma C_\sigma F_\sigma U_\sigma ^{-1/2} +
\calO C_\sigma Q([\tilde{H}_\sigma,F_\sigma])\\
{\partial {\cal L}_{LSDA} \over \partial
\phi^\dag} & = & -{1\over 2}\Obar \J n + {1\over 8\pi}L\phi\,.
\end{eqnarray*}}}
\end{center}

\subsection{Self-interaction correction}

The LDA and LSDA exchange-correlation functionals suffer from
self-interaction errors: the functionals do not correctly subtract
away the interaction of an electron with its own Hartree field when
the electron density is not uniform.  Perdew and Zunger
\cite{PerdewZunger} proposed a scheme to correct for these errors (the
SIC-LDA which we simply refer to as SIC below).

The idea is to subtract the individual electrostatic and
exchange-correlation contributions due to the density
$n_i(r)=\|\psi_i(r)\|^2$ of each quantum state $\psi_i(r)$ from the
LDA functional.  This procedure has the virtue of yielding the correct
result for a one-electron system as well as correcting for the Hartree
self-interaction exactly.  In terms of our formalism, we define the
density matrix $P_i$ and electron density $n_i$ for the state $i$ and
relate them to the total density matrix $P$ and total electron density
$n$ through
\[
P_i = Ce_if_ie_i^\dag C^\dag\ ,\ P = \sum_i P_i\ \ ; \ \ n_i =
\mbox{diag}(\I P_i\Idag)\ ,\ n = \sum_i n_i\ ,
\]
where $e_i$ is the column vector with unity in the $i$th entry and
zero elsewhere.  In addition to the total Hartree field $\phi$, we
also introduce Hartree fields $\phi_i$ for each state $i$, and the SIC
Lagrangian takes the form
\begin{eqnarray*}
{\cal L}_{SIC}(C,\{\phi_i\}) & = & -{1\over
2}\mbox{Tr}\left(FC^\dag LC\right) +
\left(\J n\right)^\dag
\left[V_{ion} + \calO\J\exc(n)
-  \Obar\phi\right] + {1\over 8\pi}\phi^\dag L\phi\\
& & -\sum_i (\J n_i)^\dag \left[ \calO\J\exc(n_i) - 
\Obar\phi_i \right] - {1\over 8\pi}\sum_i \phi_i^\dag L \phi_i
\,.
\end{eqnarray*}

Setting the variation with respect to $\phi_i$ and $\phi$ to zero
(cf. Section~\ref{sec:KohnShamPoisson}) results in the Poisson
equations
\[
L\phi = 4\pi \Obar\J n\ , \ L\phi_i = 4\pi
\Obar\J n_i\, .
\]
Substituting these solutions into the SIC Lagrangian yields the
familiar SIC energy
\begin{eqnarray*}
E_{SIC}(C) & = & -{1\over 2}\mbox{Tr}\left(FC^\dag LC\right) +
\left(\J n\right)^\dag
\left[V_{ion} + \calO\J\exc(n)
-  {1\over 2}\Obar(4\pi L^{-1}\Obar\J n)\right] \\
& & -\sum_i (\J n_i)^\dag \left[ \calO\J\exc(n_i) - 
{1\over 2}\Obar(4\pi L^{-1}\Obar\J n_i) \right] .
\end{eqnarray*}

The derivatives of the SIC Lagrangian with respect to the density
matrices $P_i$ generate state-dependent Hamiltonians $H_i$ and
state-dependent potentials $V_i$ given by
\begin{eqnarray*}
H_i & = & -{1\over 2}L + \Idag\left[\mbox{Diag}(V_{sp}-V_i)\right]\I\ ,\\
V_i & = & \Jdag\calO\J\exc(n_i) + 
\left[\mbox{Diag }\exc'(n_i)\right]\Jdag\calO\J n_i -
\Jdag\Obar\phi_i\ ,
\end{eqnarray*}
where the state-independent potential $V_{sp}$ is that of
Eq.~(\ref{eq:Hspdef}).  The derivation of the expression for the
derivative of the Lagrangian with respect to $Y$ follows precisely the
same steps as in Section~\ref{sec:Yderiv}, and the final form is
\begin{eqnarray*}
\left({\partial {\cal L}_{SIC} \over \partial Y^\dag}\right) & = &
\left(I-\calO CC^\dag\right)\left(\sum_i H_iCe_if_ie_i^\dag\right)
U^{-1/2} + \calO CQ\left(\sum_i
\left[\tilde{H}_i,e_if_ie_i^\dag\right]\right),\nonumber\\
\tilde{H}_i & = & C^\dag H_i C\, .
\end{eqnarray*}
An examination of this form shows that to compute the derivative, each
Hamiltonian $H_i$ need only be applied to the $i$th column of $C$ (as
the product $Ce_i$ occurs in all places), so that computation of the
derivative is only slightly more demanding than the corresponding LDA
derivative.

The above results for the derivative are a generalization of those in
\cite{GoedeckerUmigar}.  Those authors, however, work in the
traditional real-space representation (where necessarily all the sums
over indices appear) and, at each step of the minimization,
orthonormalize their wave functions, so that their expressions are a
special case of ours when $U=I$.

The summary of the SIC Lagrangian and derivatives follows.
\begin{center}
\framebox[\textwidth]{\parbox{\textwidth}{
\begin{eqnarray*}
{\cal L}_{SIC}(Y,\phi,\{\phi_i\})
& = & -{1\over 2}\mbox{Tr}\left(FC^\dag LC\right) +
\left(\J n\right)^\dag
\left[V_{ion} + \calO\J\exc(n)
-  \Obar\phi\right]\\
& & -\sum_i (\J n_i)^\dag \left[ \calO\J \exc(n_i) - 
\Obar\phi_i \right] + {1\over 8\pi}\phi^\dag L\phi
- {1\over 8\pi}\sum_i \phi_i^\dag L \phi_i\,,\\
{\partial {\cal L}_{SIC} \over \partial Y^\dag} & = &
\left(I-\calO CC^\dag\right)\left(\sum_i H_iCe_if_ie_i^\dag\right)
U^{-1/2} + \calO CQ\left(\sum_i
\left[\tilde{H}_i,e_if_ie_i^\dag\right]\right)\,,\\
{\partial {\cal L}_{SIC} \over \partial
\phi^\dag} & = & -{1\over 2}\Obar\J n + {1\over
8\pi}L\phi\ \ \ ,\ \ \ {\partial {\cal L}_{SIC} \over \partial \phi_i^\dag}
 = {1\over 2}\Obar\J n_i - {1\over
8\pi}L\phi_i\,.
\end{eqnarray*}}}
\end{center}

\subsection{Band-structure and fixed Hamiltonian calculations}

Very often, we aim to find a set of quantum states $\psiset$ that are
the lowest energy eigenstates of a fixed Hamiltonian.  One case where
this occurs is in the calculation of band structures for solids within
DFT, where one has already found the stationary point of the
Lagrangian and the optimal electron density $n(r)$.  One then aims to
explore the band structure for various values of $k$-vectors.  (See
Appendix~\ref{appendix:kpts} for a full discussion of $k$-points.)
This requires finding the lowest energy eigenstates of the
Hamiltonian.  The problem is the same as a tight-binding calculation
in the sense that the Hamiltonian is fixed and the electronic energy
of the system is sought after, i.e. the minimum of the expectation
value of the Hamiltonian among an orthonormal set of states.  In both
cases, the approach described below is most useful when the number of
basis functions is much larger than the number of states $\psiset$ so
that direct diagonalization of the Hamiltonian is computationally
prohibitive.

In such cases, we have a Hamiltonian $H$, and we expand our wave
functions as shown in Eq.~(\ref{eq:defCphi}).  We must minimize the
energy $E$
\[
E = \mbox{Tr}(C^\dag HC)\,.
\]
We introduce unconstrained variables $Y$ in the same way as before
(Eq.~(\ref{eq:Udef}) and onwards).  The variation of the energy is
given by
\[
dE = \mbox{Tr}(H\,d[YU^{-1}Y^\dag ]) = 2\,\mbox{Re}\,\mbox{Tr}\left[dY^\dag
\left({\partial E \over \partial Y^\dag}\right)\right]\,,
\]
where the derivative of $E$ is
\[
\left({\partial E \over \partial Y^\dag}\right)
 = (I-\calO CC^\dag )HCU^{-1/2}\,.
\]
When we are at the minimum of $E$, we have an orthonormal set $C$ that
spans the subspace of the lowest-energy eigenstates of $H$.  The
minimum value of $E$ is the electronic energy for the case of a
tight-binding Hamiltonian.  If the energy eigenvalues and eigenvectors
are desired, we diagonalize the subspace Hamiltonian $\tilde{H} =
C^\dag HC$ to obtain the eigenvalues $\epsilon$.  We then use the
unitary matrix $S$ which diagonalizes $\tilde{H}$,
$\tilde{H}=S(\mbox{Diag }\epsilon)S^\dag$, to find the expansion
coefficients for the eigenstates, given by the product $CS$.  The
summary of key equations follows.
\begin{center}
\framebox[\textwidth]{\parbox{\textwidth}{
\begin{eqnarray*}
E(Y) & = & \mbox{Tr}(C^\dag HC)\,,\\
{\partial E \over \partial Y^\dag} & = & (I-\calO CC^\dag )HCU^{-1/2}\,.
\end{eqnarray*}}}
\end{center}

\subsection{Unoccupied states}

A slightly more complex variant of the previous problem arises when we
have converged a calculation, found the orthonormal states $C$
spanning the occupied subspace, and are interested in finding the
eigenvalues and eigenstates for the low-lying unoccupied states.  For
example, let us say that we have converged a calculation in an
insulator or semi-conductor, where the occupied space specifies the
valence band.  We wish to find the low-lying conduction states in
order to study the band structure and band-gap of the material.

Thus, we start with a fixed Hamiltonian $H$ and a fixed set of
occupied states $C$.  We aim to find a set of orthonormal unoccupied
states $D$ that are orthogonal to $C$ and which also minimize the
expectation of the Hamiltonian.  Specifically, we wish to minimize
$E=\mbox{Tr}(D^\dag HD)$ under the orthogonality constraint
$D^\dag\calO C=0$.

We introduce a set of unconstrained states $Z$.  We project out the
part of $Z$ lying in the occupied subspace by using the projection
operator $\bar\rho^\dag = I-CC^\dag\calO$ of
Section~\ref{sec:KohnShamPoisson},
\[
D = \bar\rho^\dag ZX^{-1/2}\ \ \ \mbox{where}\ \ \ X = (\bar\rho^\dag
Z)^\dag\calO(\bar\rho^\dag Z).
\]
Then, following the results of the previous section, the differential
of $E$ is given by
\[
dE = \mbox{Tr}(H\,d[\bar\rho^\dag ZX^{-1}\bar\rho Z^\dag ]) =
2\,\mbox{Re}\,\mbox{Tr}\left[dZ^\dag \left({\partial E \over \partial
Z^\dag}\right)\right]\ ,
\]
where the derivative of $E$ with respect to $Z$ is given by
\[
\left({\partial E \over \partial Z^\dag}\right) = (I-\calO CC^\dag)
(I-\calO DD^\dag)HDX^{-1/2}.
\]
As expected, the derivative has two projection operators: $\bar\rho =
I-\calO CC^\dag$, which projects out the component lying in the
occupied subspace, and $(I-\calO DD^\dag)$, which projects out the
component lying in the portion of the unoccupied subspace under
consideration.  Minimization of $E$ leads to a set of states $D$ that
span the lowest-lying unoccupied states.  At the minimum, the
resulting unoccupied subspace Hamiltonian $\bar{H}=D^\dag HD$ can be
diagonalized to obtain the desired eigenvalues and eigenstates.

The energy and its gradient are summarized by
\begin{center}
\framebox[\textwidth]{\parbox{\textwidth}{
\begin{eqnarray*}
E(Z) & = & \mbox{Tr}(D^\dag HD)\,,\\
{\partial E \over \partial Z^\dag} & = &
(I-\calO CC^\dag)(I-\calO DD^\dag)HDX^{-1/2}\,,\\
D & = & \bar\rho^\dag ZX^{-1/2}\ \ \ ,\ \ \ X = (\bar\rho^\dag Z)^\dag
\calO(\bar\rho^\dag Z)\ \ \ ,\ \ \bar\rho = I-\calO CC^\dag\,.
\end{eqnarray*}}}
\end{center}

\subsection{Variational density-functional perturbation theory}

In this final application, we consider perturbation theory within a
single-particle formalism, which is required to compute response
functions.  Specifically, we consider the important case of linear
response, which was first dealt with in \cite{Gonze}.  We imagine that
we have converged the calculation of the zeroth-order
(i.e. unperturbed) configuration and have found the zeroth-order wave
functions $C_0$ for our problem.  We now wish to find the first-order
changes of the wave functions, $C_1$, due to an external perturbation
to the system.  Depending on the type of perturbation applied, the
variation $C_1$ allows for the calculation of the corresponding
response functions.  For example, the displacement of atoms along a
phonon mode allows for the computation of the dynamical matrix for that
mode whereas perturbations due to an external electrostatic field
allow for calculation of the dielectric tensor.

Regardless of the physical situation, all perturbations enter as a
change in the ionic (or external) potential $V_{ion}$ which drives the
electronic system.  Letting $\lambda$ be the perturbation parameter,
we expand any physical quantity $A$ in powers of $\lambda$ and let
$A_n$ be the coefficient of $\lambda^n$ in the expansion.  A few
examples follow
\begin{eqnarray*}
V_{ion} & = & V_{ion,0} + \lambda V_{ion,1} + \lambda^2 V_{ion,2} +
\cdots\\
C & = & C_0 + \lambda C_1 + \lambda^2 C_2 + \cdots\\
n & = & n_0 + \lambda n_1 + \lambda^2 n_2 + \cdots
\end{eqnarray*}
As is well known from perturbation theory, the first order change
$V_{ion,1}$ determines the first order shift of the wave functions
$C_1$.

The work of \cite{Gonze} shows that $C_1$ can be obtained via the
constrained minimization of an auxiliary quadratic functional of
$C_1$.  In our matrix-based notation, for the case of constant
fillings (taken to be unity) and the LDA approximation, this quadratic
functional is given by
\begin{eqnarray*}
E(C_1) & = & \mbox{Tr}\left\{C_1^\dag H_0C_1 - C_1^\dag\calO
C_1[\mbox{Diag }\epsilon_0]\right\}\\ & & + (\J
n_1)^\dag\left\{V_{ion,1} + \calO\J \left[\mbox{Diag }
a(n_0)\right]n_1 - {1\over 2}\Obar\phi_1
\right\}+E_{nonvar}\,.
\end{eqnarray*}
The energy $E_{nonvar}$ contains terms that depend only on $C_0$ or
the Ewald sum over atomic positions and need not concern us any
further.  The zeroth-order Hamiltonian $H_0 = -{1\over 2}L +
\Idag[\mbox{Diag }V_0]\I$ is the same as that of Eq.~(\ref{eq:Hspdef})
where we have simply renamed the zeroth-order single-particle
potential to $V_0$.  The diagonal matrix $\mbox{Diag }\epsilon_0$
holds the eigenvalues of the zeroth-order Hamiltonian.  The vector
$a(n_0)$ is found by evaluating the function $a(n)={d^2 \over
dn^2}\left(n\exc(n)\right)$ on the zeroth-order electron density $n_0$
over the real-space grid.  The vector $n_1$, the first order shift of
the electron density, is given by
\[
n_1 = \mbox{2 Re diag}\left( \I C_0 C_1^\dag \Idag \right)\,.
\]
The first order change of the Hartree potential $\phi_1$ is the solution
of the Poisson equation $\phi_1 = 4\pi L^{-1}\Obar\J n_1$.

Given the quadratic nature of $E(C_1)$, its differential with respect
to $C_1$ follows immediately and is given by
\[
dE = \mbox{2 Re Tr}\left\{ dC_1^\dag \left(H_0C_1 -
\calO C_1[\mbox{Diag }\epsilon_0]
+ \Idag\left[\mbox{Diag }V_1\right]\I C_0\right)\right\}\,,
\]
where the first-order single-particle potential $V_1$ is given by
\[
V_1 = \Jdag V_{ion,1} +
\Jdag\calO\J \left[\mbox{Diag }a(n_0)\right]n_1 +
\left[\mbox{Diag }a(n_0)\right]\Jdag\calO\J n_1 -
\Jdag\Obar\phi_1\,.
\]

The constraint to be obeyed during the minimization is that the
first-order shifts $C_1$ be orthonormal to the zeroth-order wave
functions $C_0$,
\[
C_1^\dag \calO C_0 = I\,.
\]
This constraint is easily handled in the manner of the previous
section by using a projection operator.  We introduce an unconstrained
set of wave functions $Y_1$ from which we project out the part laying
in the space spanned by $C_0$,
\[
C_1 = (I - C_0 C_0^\dag \calO)Y_1\,.
\]
Based on this relation, we find the gradient of $E$ with respect to
$Y_1$
\begin{eqnarray*}
dE & = & \mbox{2 Re Tr}\left\{dY_1^\dag \left({\partial E \over \partial
Y_1^\dag}\right)\right\}\mbox{  where}\\
\left({\partial E \over \partial Y_1^\dag}\right) & = &
(I-\calO C_0 C_0^\dag)\left\{H_0C_1 - \calO C_1[\mbox{Diag}\epsilon_0]
+ \Idag \left[\mbox{Diag }V_1\right]\I C_0\right\}\,.
\end{eqnarray*}

Finally, we can convert the energy function into a Lagrangian by
letting $\phi_1$ be a free variable and by adding the appropriate
Hartree self-energy and coupling to $n_1$.  We arrive at the
summarized expressions
\begin{center}
\framebox[\textwidth]{\parbox{\textwidth}{
\begin{eqnarray*}
{\cal L}(Y_1,\phi_1) & = &
\mbox{Tr}\left\{C_1^\dag H_0C_1 - C_1^\dag\calO
C_1[\mbox{Diag }\epsilon_0]\right\} + E_{nonvar}\\
& & + (\J n_1)^\dag\left\{V_{ion,1} + \calO\J \left[\mbox{Diag }
a(n_0)\right]n_1 - \Obar\phi_1\right\} +
{1\over 8\pi}\phi_1^\dag L\phi_1\,,\\
{\partial {\cal L} \over \partial Y_1^\dag} & = &
(I-\calO C_0 C_0^\dag)\left\{H_0C_1 - \calO C_1[\mbox{Diag }\epsilon_0] +
\Idag \left[\mbox{Diag }V_1\right]\I C_0\right\}\,,\\
{\partial {\cal L} \over \partial \phi_1^\dag} & = &
-{1\over 2}\Obar\J n_1 + {1\over
8\pi}L\phi_1\,.
\end{eqnarray*}}}
\end{center}

\section{Minimization algorithms}
\label{sec:minalgs}

In this section, we show how the DFT++ formalism can succinctly
specify the algorithm which finds the stationary point of the
Lagrangian or energy function (derived in the previous sections).
Such an algorithm only requires the value and derivative of the
objective function, which is the reason that we have repeatedly
emphasized the importance of these two quantities in our analysis
above.  Once we choose a minimization algorithm, we need only
``plug in'' the objective function and its derivative into the
appropriate slots.  Furthermore, since the DFT++ formalism is compact
and at the same time explicit, once we specify the operations that
must be performed for a given algorithm in our notation, the
transition to coding on a computer is trivial: the formulae translate
directly into computer code (as shown in
Section~\ref{sec:imploptpara}).

\begin{figure}[t!b!]
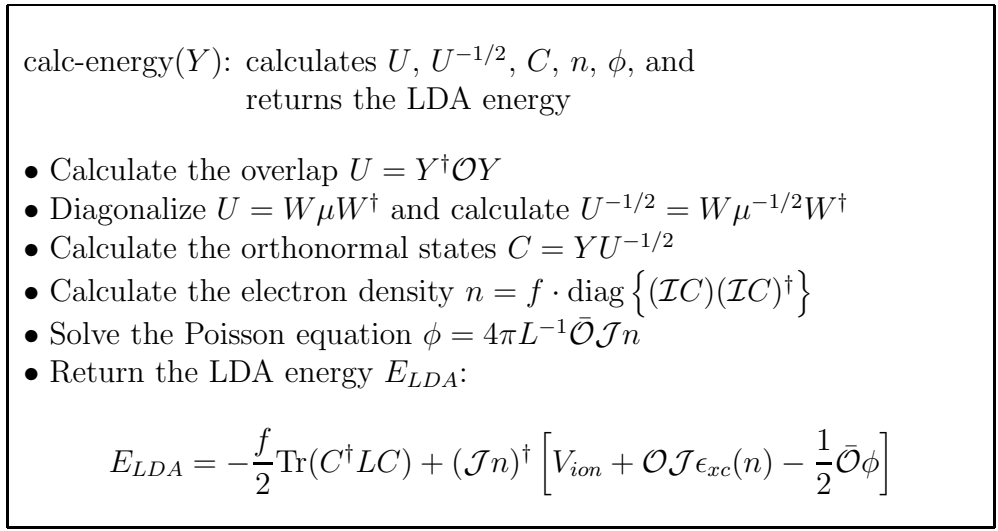

\begin{center}
\begin{tabular}{|l|}
\hline
\\
calc-energy($Y$):  \parbox[t]{3in}{
calculates $U$, $U^{-1/2}$, $C$, $n$, $\phi$, and\\
returns the LDA energy}\\
\\
$\bullet$ Calculate the overlap $U=Y^\dag\calO Y$\\
$\bullet$ Diagonalize $U=W\mu W^\dag$ and calculate 
$U^{-1/2}=W\mu^{-1/2} W^\dag$\\
$\bullet$ Calculate the orthonormal states $C=YU^{-1/2}$\\
$\bullet$ Calculate the electron density
$n=f\cdot\mbox{diag}\left\{(\I C)(\I C)^\dag\right\}$\\
$\bullet$ Solve the Poisson equation $\phi = 4\pi L^{-1}\Obar\J n$\\
$\bullet$ Return the LDA energy $\ELDA$:\\
\parbox{5in}{\[\ELDA = -{f\over 2}\mbox{Tr}(C^\dag
LC) + (\J n)^\dag \left[V_{ion} + \calO\J \exc(n) -
{1\over 2}\Obar \phi\right]\]}\\
\hline
\end{tabular}
\caption{\label{fig:calcener}LDA energy routine (DFT++ formalism)}
\end{center}
\end{figure}

Specifically, we aim to find the stationary point of the Lagrangian of
Eq.~(\ref{eq:lagrexpr1}) with respect to $Y$ and $\phi$ and possibly
$B$.  A direct search for the stationary point is possible, and at the
saddle point, both the Kohn-Sham and Poisson equations hold true
simultaneously.  This highly effective strategy has been followed in
\cite{torkelarias}.  Alternatively, other approaches to reach a
solution of these equations through self-consistent iteration and use
of Broyden-like schemes \cite{Broyden} may be considered.

However, in order to make direct contact with DFT calculations within the
traditional minimization context \cite{RMP}, and to keep our
presentation as simple as possible, we choose instead to solve the
Poisson equation (Eq.~(\ref{eq:Poissonexpr2})) for the optimal $\phi$
at each iteration of the minimization algorithm.  For cases where
$L^{-1}$ is easy to compute (e.g. the plane-wave basis where $L$ is
diagonal), we may compute the solution $\phi$ directly from
Eq.~(\ref{eq:Poissonsolution}).  Otherwise, the straightforward
approach of maximizing the quadratic functional ${\cal G}(\phi) =
(1/8\pi)\phi^\dag L\phi - (\J n)^\dag\Obar\phi$ via conjugate
gradients (or some other method) achieves the same goal.  For certain
basis sets, multigrid methods or other specialized techniques are
possibilities as well \cite{torkelarias,multigrid,fastgauss}.  Once
the Poisson equation has been solved, the remaining free variable is
the matrix of coefficients $Y$, and the aim of the calculation is to
minimize the energy $\ELDA$ of Eq.~(\ref{eq:ELDAexpr}) with respect to
$Y$.

As shown in Section~\ref{sec:lagr}, the value and $Y$-derivatives of
the Lagrangian $\LLDA$ and energy $\ELDA$ are identical if we evaluate
the Lagrangian-based expressions using the Hartree potential $\phi$
which is the solution of Poisson's equation.  Therefore, in our
algorithms below, we can use expressions derived for derivatives of
the Lagrangian when dealing with the energy.

\subsection{Semiconducting and insulating systems}
\label{sec:pcg}

\begin{figure}[b!t!]
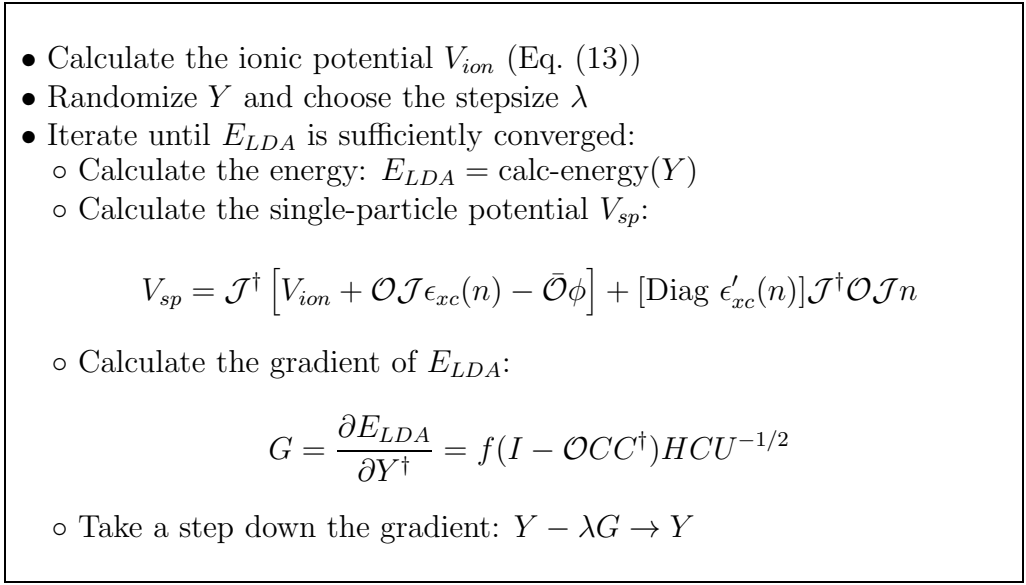

\begin{center}
\begin{tabular}{|l|}
\hline \\ $\bullet$ Calculate the ionic potential $V_{ion}$
(Eq.~(\ref{eq:Viondef}))\\ $\bullet$ Randomize $Y$ and choose the
stepsize $\lambda$\\ $\bullet$ Iterate until $\ELDA$ is sufficiently
converged:\\ \ \ \ \parbox{5in}{ $\circ$ Calculate the energy: $\ELDA
= \mbox{calc-energy}(Y)$\\ $\circ$ Calculate the single-particle
potential $V_{sp}$:\\
\parbox{5in}{\[V_{sp} = \Jdag\left[V_{ion} +
\calO\J \exc(n) - \Obar\phi\right]
+ [\mbox{Diag }\excprime(n)]\Jdag\calO\J n\]}\\
$\circ$ Calculate the gradient of $\ELDA$:\\
\parbox{5in}{\[G = {\partial \ELDA \over \partial Y^\dag}
= f(I-\calO CC^\dag )HCU^{-1/2}\]}\\
$\circ$ Take a step down the gradient: $Y-\lambda G\rightarrow Y$
}\\
\\
\hline
\end{tabular}
\caption{\label{fig:sdalgorithm}Steepest descent algorithm}
\end{center}
\end{figure}

Consider the case of a semiconducting system with a large unit cell.
The fillings are constant, $F=fI$, and we will use a single $k$-point
at $k=0$ (as appropriate for a large cell).  The simplest algorithm
for minimizing the energy is the steepest descent method: we update
$Y$ by shifting along the negative gradient of the energy with respect
to $Y$, scaled by a fixed multiplicative factor.  As a first
organizational step, we introduce the function calc-energy($Y$), whose
code we display in Figure~\ref{fig:calcener}.  Given $Y$, this
function computes the overlaps $U$ and $U^{-1/2}$, the orthonormalized
$C$, the electron density $n$, the solution to Poisson's equation
$\phi$, and returns the LDA energy.  Figure~\ref{fig:sdalgorithm}
displays the steepest descent algorithm as it appears in the DFT++
language.

We would like to emphasize a number of points regarding this code.
First, the algorithm optimizes all the wave functions (i.e. the entire
matrix $Y$) at once, leading to a very effective minimization and a
complete avoidance of charge creep \cite{ACprl} or other
instabilities.  Second, the code is independent of the
basis set used: the basis-dependent operators $L$, $\calO$, etc., are
coded as low-level functions that need to be written only once.  The
choice of physical system and minimization algorithm is a high-level
matter that is completely decoupled form such details.  Third, the
figure shows {\em all} the operations required for the entire
minimization loop.  That this is possible is grace to the succinct
matrix formalism.

\begin{figure}[t!b!]
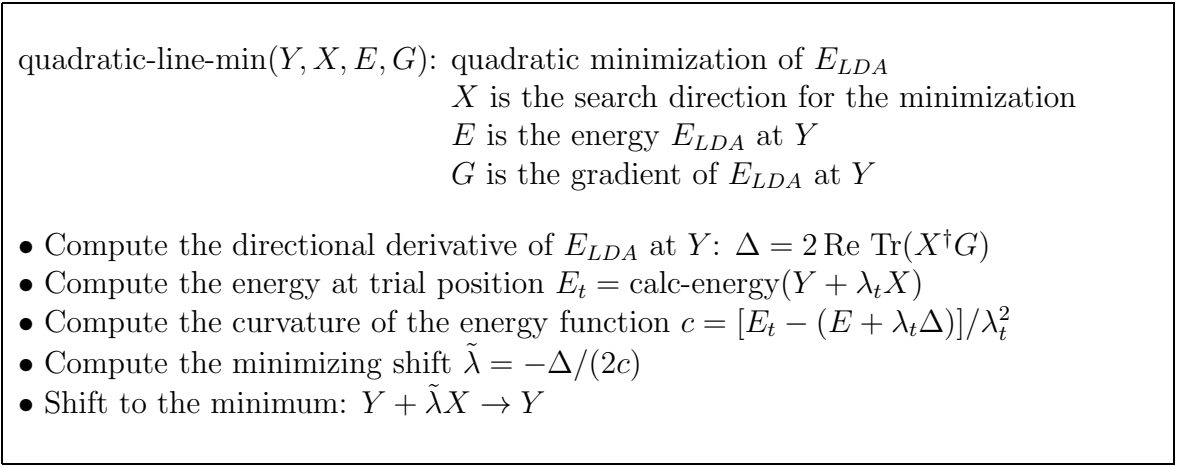

\begin{center}
\begin{tabular}{|l|}
\hline
\\
quadratic-line-min($Y,X,E,G$): \parbox[t]{3.7in}{quadratic minimization of
$\ELDA$\\
$X$ is the search direction for the minimization\\
$E$ is the energy $\ELDA$ at $Y$\\
$G$ is the gradient of $\ELDA$ at $Y$}\\
\\
$\bullet$ Compute the directional derivative of $\ELDA$ at $Y$:  $\Delta =
2\,\mbox{Re Tr}(X^\dag G)$\\
$\bullet$ Compute the energy at trial position
$E_t = \mbox{calc-energy}(Y+\lambda_t X)$\\
$\bullet$ Compute the curvature of the energy function $c = [E_t -
(E+\lambda_t\Delta)]/\lambda_t^2$\\
$\bullet$ Compute the minimizing shift $\tilde\lambda = -\Delta/(2c)$\\
$\bullet$ Shift to the minimum:  $Y+\tilde\lambda X \rightarrow Y$\\
\\
\hline
\end{tabular}
\caption{\label{fig:linmin}Quadratic line minimizer}
\end{center}
\end{figure}

The only part of the algorithm of Figure \ref{fig:sdalgorithm} that is
specific to the steepest descent method is the last operation where $Y$ is
updated.  To generalize to a preconditioned conjugate-gradient
algorithm is quite straightforward, and we specify the necessary
changes below.

Conjugate-gradient algorithms require line minimization of the
objective function along a specified direction, i.e. an algorithm is
needed that minimizes $E(\lambda)\equiv \ELDA(Y+\lambda X)$ with
respect to the real number $\lambda$ for a fixed search vector $X$.
The subject of line minimization is rich, and an abundance of
algorithms with varying degrees of complexity exist in the
literature. (For a brief introduction see \cite{NumRecip}.)  However,
for typical DFT calculations, most of the effort for the calculation
is spent in the quadratic basin close to the minimum.  Thus, a simple
and efficient line-minimizer that exploits this quadratic behavior
should be sufficient, and we have found this to be the case in our
work.

Our line-minimizer takes the current value of the energy and its
derivative as well as the value of the energy at the shifted trial
configuration $Y+\lambda_tX$ (where $\lambda_t$ is a trial step-size)
to achieve the quadratic fit $E(\lambda)\approx E + \Delta\lambda +
c\lambda^2$.  Here, $\Delta$ is the directional derivative of $E$ with
respect to $\lambda$, and $c$ is the curvature of $E$ with respect to
$\lambda$.  The line-minimizer then moves to the minimum of this
quadratic fit located at $\tilde\lambda = -\Delta/(2c)$. Figure
\ref{fig:linmin} shows the code for this line-minimizer.

\begin{figure}[b!t!]
\begin{center}
\begin{tabular}{|l|}
\hline
\\
$\bullet$ Calculate the ionic potential $V_{ion}$\\
$\bullet$ Randomize $Y_0$\\
$\bullet$ For $j$=0, 1, 2, ...\\
\ \ \ \parbox{5in}{
$\circ$ Let $E_j = \mbox{calc-energy}(Y_j)$\\
$\circ$ Calculate the single-particle potential $V_{sp}$\\
$\circ$ Calculate the gradient: $G_j = (\partial
\ELDA/\partial Y^\dag)|_{Y_j}$\\
$\circ$ Apply the preconditioner: ${\Gamma}_j = K(G_j)$\\
$\circ$ If $j>0$ then set $\alpha_j = \mbox{Re Tr}(G_j^\dag {\Gamma}_j)/
\mbox{Re Tr}(G_{j-1}^\dag\,{\Gamma}_{j-1})$; otherwise $\alpha_j=0$.\\
$\circ$ Compute the search direction $X_j = {\Gamma}_j + \alpha_j X_{j-1}$\\
$\circ$ Call quadratic-line-min($Y_j,X_j,E_j,G_j$)\\
}\\
\hline
\end{tabular}
\caption{\label{fig:cgalgorithm}Preconditioned conjugate-gradient
algorithm}
\end{center}
\end{figure}

Using this line minimizer, we present the entire preconditioned
conjugate-gradient algorithm in Figure \ref{fig:cgalgorithm}.  Note
that we have omitted some of the formulae which are identical to those
of Figure \ref{fig:sdalgorithm}. A simple diagonal preconditioning, as
described in \cite{RMP}, is quite effective for plane-wave basis sets,
and the operator $K$ is the preconditioner in the algorithm of the
figure.

\subsection{Metallic and high-temperature systems}
\label{sec:subspacerot}

While the degrees of freedom in the variable $Y$ are sufficient to
describe any electronic system, the convergence of minimization
algorithms can be hampered by ill-conditioning of the physical system.
A standard case of such a problem is when the Fermi-Dirac fillings
$f_i$ are not constant and some fillings are very small, a situation
encountered when studying metals or high-temperature insulators.

One type of ill-conditioning that arises due to states with small
fillings, $f_i\ll 1$, stems from the fact that changes in such states
do not affect the value of the energy $\ELDA$ very much when compared
to the states with large fillings, $f_i\sim 1$.  Thus modes associated
with the small-filling states are ``soft'' and we have an
ill-conditioned problem.  The solution to this problem, however, is
rather straightforward and involves scaling the derivative of $\LLDA$
with respect to the state $\psi_i$ by $1/f_i$.

Much harder to deal with are soft modes due to subspace rotations
which were introduced in Section~\ref{sec:orthonorm}.  As we saw
there, the unitary transformations $V$, which generate the rotations,
cancel out completely from the expression for the density matrix $P$
in the case of constant Fermi fillings, $F=fI$.  Since the entire
energy can be computed from $P$ alone, the energy will not depend on
$V$.  Hence we have found that the unitary transformations $V$ are an
exact symmetry of a system with constant fillings.

However, once we introduce variations in the fillings, the symmetry is
broken.  Now both the density matrix and the energy change when $V$
mixes states with different fillings.  If the difference in fillings
between the mixed states is small, a case typically encountered in a
real system, the mixing produces small changes in the energy.  Again,
we have soft modes, this time due to the breaking of the unitary
subspace-rotation symmetry.

The idea of using subspace rotations was first suggested in
\cite{ACprl}.  Its use as a cure for the ill-conditioning described
above was discussed and demonstrated convincingly in \cite{MVP}.  The
strategy is first to minimize the objective function over $B$ (since
$B$ parameterizes the rotation $V$) and only then perform minimization
on the wave functions $Y$.  By ensuring that we are always at the
minimum with respect to $B$, we automatically set the derivative of
$\ELDA$ with respect to subspace-rotations to zero.  When this is
true, changing $Y$ can not (to first order) give rise to contributions
due to the soft modes, and we have eliminated the ill-conditioning.

In practice, we have found it unnecessary for our calculations to be
at the absolute minimum with respect to $B$: being close to the
minimum is sufficiently beneficial computationally.  In our algorithm,
we take steps along both $Y$ and $B$ simultaneously but ensure that
our step-size in $B$ is much larger than that in $Y$.  As the
minimization proceeds, this choice automatically drives the system to
stay close to the minimum along $B$ at all times.

Our simpler procedure has been found as effective as the original
strategy of \cite{MVP} and translates into using the following search
direction $X$ in the space $(Y,B)$
\[
X = \left({\partial \LLDA  \over \partial Y^\dag}\ ,\ \eta\cdot
{\partial \LLDA  \over \partial B}\right)
\]
where $\eta$ is a numerical scaling factor.  We have found $\eta \sim
30$ to be a good choice for efficient minimization while avoiding the
ill-conditioning mentioned above.

\begin{figure}[t!b!]
\begin{center}
\resizebox{4.5in}{!}{\includegraphics{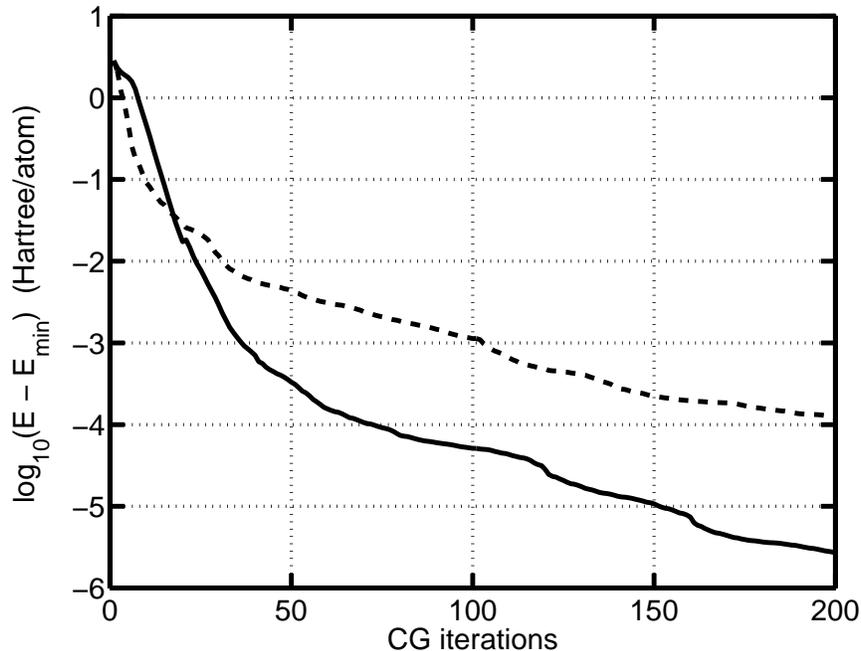}}
\caption[Effect of subspace rotation on
convergence]{\label{fig:subspace}Effect of subspace rotation on
convergence -- Convergence of conjugate-gradient minimization with
(solid) and without (dashed) the use of subspace rotations for the
case of a metallic system.  Both minimizations use the same random
wave functions as their starting points.  The horizontal axis is the
number of conjugate-gradient iterations and the vertical axis is the
energy per atom above the minimum energy in Hartree per atom in
logarithmic units.  See the text for further details.}
\end{center}
\end{figure}

As a practical showcase of the improvement in convergence in a
metallic system, we study the convergence rate for the
conjugate-gradient minimization of the energy of bulk molybdenum.  We
study the bcc cubic unit cell containing two Mo atoms.  We use a
plane-wave basis set (details of implementation in
Appendix~\ref{appendix:implementpw}) with an electronic cutoff of 22.5
Hartree (45 Ryd), and a 4$\times$4$\times$4 cubic $k$-point mesh to
sample to Brillouin zone.  The electronic temperature used is
$k_BT$=0.0037 Hartree (0.1 eV), the Fermi fillings are recomputed
every twenty conjugate-gradient steps based on the eigenvalues of the
subspace Hamiltonian from the previous iteration, and a value of
$\eta=30$ is used to calculate the search direction.  Non-local
pseudopotentials of the Kleinmann-Bylander form \cite{KB} are used
with $p$ and $d$ projectors.  Figure \ref{fig:subspace} presents a
plot of the convergence of the energy per atom to its minimum value
(as determined by a run with many more iterations than shown in the
figure).  We compare minimization with and without the use of subspace
rotation variables, and both minimizations are started with the same
initial random wave functions.  The extra cost required for the use of
subspace rotations was very small in this case, the increase in the
time spent per iteration being less than two percent.  As we can see,
the use of subspace rotations dramatically improves the convergence
rate.

\section{Implementation, optimization, and
parallelization}
\label{sec:imploptpara}

In this final section we address how the DFT++ formalism can be easily
and effectively implemented on a computer, and what steps must be
taken to ensure efficient optimization and parallelization of the
computations.  As is clear from the previous sections, the DFT++
formalism is firmly based on linear algebra.  The objects appearing in
the formalism are vectors and matrices.  The computations performed on
these objects are matrix addition and multiplication and the
application of linear operators.  An important benefit is that
linear-algebraic products involve matrix-matrix multiplications
(i.e. BLAS3 operations), which are precisely those operations that
achieve the highest performance.

We use an object-oriented approach and the C++ programming language to
render the implementation and coding as easy as possible.  In
addition, object-oriented programming introduces modularity and
localization of computational kernels allowing for effective
optimization and parallelization.  In the sections below, we present
outlines of our implementation, optimization, and parallelization
strategies.

\subsection{Object-oriented implementation in C++}
\label{sec:cppimplem}

In our work, we have found that an object-oriented language such as
C++ is ideal for implementing the required vectors and matrices and
for defining the operations on them in a manner that follows the DFT++
formalism as closely as possible.  The object-oriented approach
presents a number of advantages.

First, the programming task becomes highly modular: we identify the
object types needed and the operations that must be performed on them,
and we create a separate module for each object that can be tested
independently.  For example, we define the class of matrices and the
operations on them (e.g. multiplication, addition, diagonalization,
etc.), and we can test and debug this matrix module separate from any
other considerations.  Second, we gain transparency: the internal
structure or functioning of an object can be modified for improved
performance without requiring any changes to higher-level functions
that use the object.  This gives us a key feature in that the
high-level programmer creating new algorithms or testing new energy
functionals does not need to know about or interact with the
lower-level details of how the objects actually are represented or how
they function.  Third, this separation of high-level function from
low-level implementation allows for a centralization and reduction in
the number of computational kernels in the code: there can be a large
variety of high-level objects for the convenience of the programmer,
but as all the operations defined on them are similar linear-algebraic
ones, only a few actual routines must be written.  Fourth, modularity
produces shorter and more legible code.  This, combined with the
object-oriented approach, implies that the high-level computer code
will read the same as the equations of the DFT++ formalism so that
debugging will amount to checking formulae, without any interference
of cumbersome loops and indices.

To give a concrete example of what this means, consider the simplest
object in the formalism, a column vector such as the electron density
$n$.  In C++, we define an object class \texttt{vector} which will
contain an array of \texttt{complex} numbers (itself a lower-level
object).  A \texttt{vector v} has a member \texttt{v.size} specifying
the number of rows it contains as well as a pointer to the array
containing the data.  We define the action of the parenthesis so as to
allow convenient access of the \texttt{i}th element of \texttt{v} via
the construction \texttt{v(i)}.

A very common operation between two vectors $v$ and $w$ is the scalar
product $v^\dag w$.  We implement this by defining (in C++ parlance
overloading) an operator \texttt{\^{}} that takes two \texttt{vector}s
and returns a \texttt{complex} result.  The code accomplishing this is

{\tt \begin{tabbing}
\codeindent\=friend complex operator\^{}(vector \&v, vector \&w)\\
\>\{\ \ \ \=\\
\> \> complex result = 0;\\
\\
\> \> for\= (int i=0; i < v.size; i++)\\
\> \> \> result += conjugate(v(i))*w(i);\\
\\
\> \> return result;\\
\>\}
\end{tabbing}}

An example of an operator acting on vectors is the inverse transform
$\J$.  This operator can be coded as a function \texttt{J} that takes
a \texttt{vector v} as its argument and returns the \texttt{vector}
result \texttt{J(v)}.  Of course, the details of what $\J$ does
internally are basis-dependent.

Based on this definition, the evaluation of the electron-ion
interaction energy of Section~\ref{sec:eiinter}, given by the
expression $E_{e-i}=(\J n)^\dag V_{ion}$, is coded by\\ \\
\parbox{\textwidth}{ {\tt\codeindent E\_ei = J(n)\^{}V\_ion;}}\\ \\
where \texttt{V\_ion} is the vector $V_{ion}$ of
Eq.~(\ref{eq:Viondef}).

Following the same idea, we define a \texttt{matrix} class to handle
the matrices such as $U$, $W$, $\tilde{H}$, etc. that occur in the
formalism.  Physically, expansion coefficients such as $Y$ and $C$ are
collections of column vectors (a column of coefficients $C_{\alpha i}$
for each wave function $\psi_i$), and we define the class
\texttt{column\_bundle} to handle these objects.  Although
mathematically \texttt{column\_bundle}s such as $Y$ and $C$ are
matrices, the choice not to use the \texttt{matrix} class for
representing them is deliberate, as $Y$ and $C$ have a distinct use
and a special physical meaning that an arbitrary matrix will not have.
In this way, we can tailor our objects to reflect the physical content
of the information they contain.  Of course, when a matrix
multiplication is performed, such as when we evaluate the expression
$C=YU^{-1/2}$, the \texttt{column\_bundle} class and \texttt{matrix}
class call a single, low-level, optimized multiplication routine.  We
thus gain flexibility and legibility of codes on the higher levels
while avoiding an accumulation of specialized functions on the lower
levels.

\begin{figure}[t!b!]
\begin{center}
\begin{tabular}{|l|}
\hline
\parbox{\textwidth}{
{\tt\begin{tabbing}
complex calc\_energy(\=column\_bundle \&Y, column\_bundle \&C,\\
\>matrix \&U, matrix \&Umhalf,\\
\>vector \&n, vector \&phi,\\
\>double f, vector \&V\_ion)\\
\{\\
\ \ \=complex E\_LDA;\\
\\
\>U = Y\^{}O(Y);\\
\\
\>// calc\_Umhalf(U) returns U\^{}\{-1/2\} given U\\
\>Umhalf = calc\_Umhalf(U);\\
\\
\>C = Y*Umhalf;\\
\\
\>// diag\_of\_self\_product(X) returns diag(X*adjoint(X))\\
\>n = f*diag\_of\_self\_product(I(C));\\
\\
\>phi = (4.0*PI)*invL(Obar(n));\\
\\
\>E\_LDA = \=-0.5*f*Tr(C\^{}L(C)) +\\
\>\>J(n)\^{}( V\_ion + O(J(exc(n))) - 0.5*Obar(phi) );\\
\\
\>return E\_LDA;\\
\}
\end{tabbing}}}
\\
\hline
\end{tabular}
\caption[LDA energy routine (C++ implementation)]
{\label{fig:calcenergycode}LDA energy routine (C++ implementation) --
C++ code for the calc-energy($Y$) function which was outlined in
Figure \protect\ref{fig:calcener}.  When computing the electron
density \texttt{n}, we have not used the straightforward code
\texttt{diag(X*adjoint(X))} which first computes the entire matrix
\texttt{X*adjoint(X)} before extracting its diagonal and is thus
computationally wasteful.  Rather, for performance reasons, we have
written a separate function \texttt{diag\_of\_self\_product(X)} that
given a \texttt{column\_bundle X} computes only the required diagonal
portion of the product \texttt{X*adjoint(X)}.}
\end{center}
\end{figure}

As a concrete example of the power and ease of this style of
programming, consider writing the code for the function
calc-energy($Y$) of Figure~\ref{fig:calcener}.  In order to do so, we
will need a few more operators.  Following the example of the
\texttt{vector}s, we define the \texttt{\^{}} operator between two
\texttt{column\_bundle}s to handle Hermitian-conjugated
multiplications such as $Y_1^\dag Y_2$, where the result of the
product is of type \texttt{matrix}.  Next, we define the \texttt{*}
operator between a \texttt{column\_bundle} and \texttt{matrix} to
handle multiplications of the type $YU^{-1/2}$, where the result is a
\texttt{column\_bundle}.  We code the action of the basis-dependent
operators such as $\calO$ or $\I$ on a \texttt{vector} $\phi$ or a
\texttt{column\_bundle} $C$ as function calls \texttt{O(phi)} or
\texttt{I(C)}, which return the same data type as their argument.
Finally, we implement multiplication by scalars in the obvious way.
Figure \ref{fig:calcenergycode} presents the C++ code for the energy
calculation routine.  The code is almost exactly the same as the
corresponding expressions in the DFT++ formalism, making the
translation from mathematical derivation to actual coding trivial.

\subsection{Scalings for dominant DFT++ operations}
\label{sec:FLOPcount}

Before we describe our approach to optimization and parallelization,
we will work out the scalings of the floating-point operation counts
as a function of system size for the various computational operations
in the DFT++ formalism.  Thus it will be clear which optimizations and
parallelizations will increase the overall performance most
efficaciously.  Let $n$ be the number of wave functions $\psiset$ and
let $N$ be the number of basis functions $\{b_\alpha(r)\}$ in the
calculation.  A \texttt{vector} contains $N$ complex numbers, a
\texttt{matrix} is $n\times n$, and a \texttt{column\_bundle} is
$N\times n$ and is the largest data structure in the computation.
Both $n$ and $N$ are proportional to the number of atoms $n_a$ in the
simulation cell, or equivalently, to the volume of the cell.
Typically, for accurate DFT calculations, the number of basis
functions required is much larger than the number of quantum states,
$N\gg n$.

\begin{table}[t!]
\begin{center}
\begin{tabular}{|c|c|c|}
\hline
Operation & Examples & FLOP count \\
\hline
\hline
\texttt{column\_bundle} & $M=Y_1^\dag Y_2$ or & $O(n^2N)$\\
multiplications & $C=YU^{-1/2}$ &  \\
\hline
\texttt{matrix} multiplications & $U=W\mu W^\dag$ or & $O(n^3)$\\
and diagonalizations & $U^{-1/2} = W\mu^{-1/2}W^\dag$ & \\
\hline
Applying basis-dependent & $\calO Y$, $LC$, or & $O(nN)$ or\\
operators & $\I C$ & $O(nN\ln N)$\\
\hline
Calculating $n$         & &\\ 
given $\I C$ or    &
$n=\mbox{diag}\{(\I C)F(\I C)^\dag\}$ & $O(nN)$ \\
calculating the kinetic & or $\mbox{Tr}\{FC^\dag(LC)\}$ & \\
energy given $LC$       & &\\
\hline
\texttt{vector} operations &
$\exc(n)$, $(\J n)^\dag V_{ion}$, or & $O(N)$\\
& $\Obar\phi$ & \\
\hline
\end{tabular}
\end{center}
\caption[FLOP count of dominant DFT++ operations]{Floating-point
operation count for various common operations in the DFT++ formalism.
The size of the basis set is $N$ and the number of wave functions is
$n$.  Thus a \texttt{column\_bundle} is $N\times n$, a \texttt{matrix}
is $n\times n$, and a \texttt{vector} is $N$ long.}
\label{table:FLOPcount}
\end{table}

In Table \ref{table:FLOPcount} we present the floating-point operation
counts for the different operations required in the DFT++ formalism.
We note that for very large systems, the basis-set independent matrix
products dominate the overall computational workload.  However,
for medium-sized or slightly larger problems, the application of the
basis-dependent operators can play an important role as well.

For most basis sets, there are techniques available in the literature
that allow for efficient application of the basis-dependent operators
to a column vector in $O(N)$ or $O(N\ln N)$ operations.  For example,
when working with plane waves, the Fast Fourier transform \cite{fft}
is the algorithm of choice for implementing the operators $\I$, $\J$,
$\Idag$, and $\Jdag$ and allows us to affect these operators in
$O(N\ln N)$ operations.  (See Appendix~\ref{appendix:implementpw} for
the details of a plane-wave implementation.)  The operators $L$ and
$\calO$ are already diagonal in a plane-wave basis and are thus
trivial to implement.  For real-space, grid-based computations, multigrid
methods \cite{multigrid} are highly effective for inverting $L$ and
solving the Poisson equation.  Special techniques exist for
multiresolution \cite{LAE,torkelarias,BCR} and Gaussian
\cite{fastgauss} basis sets that allow for the application of the
basis-dependent operators in $O(N)$ operations as well.

\subsection{Optimization of computational kernels}
\label{sec:optim}

Due to the modularity of our object-oriented approach, the physical
nature of the problem under study is a high-level issue that is
completely independent of the functioning of the few, low-level
computational kernels which handle most of the calculations in the
code.  The aim now is to optimize these kernels to obtain maximum
computational performance.  By optimization we mean increasing
performance on a single processor.  Parallelization, by which we mean
distribution of the computational task among several processors, will
be addressed in the next section, but good parallel performance is
only possible when each processor is working most effectively.

The computationally intensive operations involved in the DFT++
formalism fall into two overall classes.  First are the application of
the basis-dependent operators such as $L$, $\calO$, $\I$, etc., whose
operation counts scale quadratically in the system size (see Table
\ref{table:FLOPcount}).  Given their basis-dependent nature, no
general recipe can be given for their optimization.  However, for many
widely used basis sets, highly efficient methods exist in the
literature which allow for the application of these operators, and we
refer the reader to the references cited in the preceding section.
For the case of plane waves, we have used the FFTW package \cite{fftw}
to affect the Fourier transformations.  This highly portable, freely
obtainable software library provides excellent per processor
performance across many platforms.

The second class of operations are the basis-independent matrix
products, and we will now consider the optimization of these
operations.  As a case study, we examine the Hermitian-conjugated
matrix product between two \texttt{column\_bundle}s, which occurs in
an expression such as $Y_1^\dag Y_2$ and which is coded using the {\tt
column\_bundle\^{}column\_bundle} operator as \texttt{Y\_1\^{}Y\_2}.
The most ``naive'' and straightforward implementation of this operator
in C++ is given by

{\tt\begin{tabbing}
\codeindent\=friend matrix operator\^{}(column\_bundle \&Y\_1, 
column\_bundle \&Y\_2)\\
\>\{\ \ \=\\
\>\>// create a matrix of size Y\_1.n\_columns x Y\_2.n\_columns\\
\>\>// to hold the result\\
\>\>matrix M(Y\_1.n\_columns,Y\_2.n\_columns);\\
\>\>int i,j,k;\\
\\
\>\>for\= (i=0; i < Y\_1.n\_columns; i++)\\
\>\>\>for\= (j=0; j < Y\_2.n\_columns; j++) \{\\
\>\>\>\>M(i,j) = 0;\\
\>\>\>\>for\= (k=0; k < Y\_1.n\_rows; k++)\\
\>\>\>\>\>M(i,j) += conjugate(Y\_1(k,i))*Y\_2(k,j);\\
\>\>\}\\
\\
\>\>return M;\\
\>\}
\end{tabbing}}

While easy to follow, this implementation is quite inefficient on
modern computer architectures for large matrix sizes because the
algorithm does not take any advantage of caching.  The access pattern
to memory is not localized, and the processor must continually read
and write data to the slower-speed main memory instead of to the much
faster (but smaller) cache memory.

One solution to this problem is to resort to vendor-specific linear
algebra packages.  For example, one can use a version of LAPACK
optimized for the computational platform at hand, and this generally
results in very good performance.  An alternative choice is to perform
the optimizations by hand.  While this second choice may sacrifice
some performance, it does ensure highly portable code, and this is the
strategy that we have followed in our work.

Our basic approach to increasing performance is to use blocking: we
partition each of the input and output matrices into blocks of
relatively small dimensions, and the matrix multiplication is
rewritten as a set of products and sums over these smaller blocks.
Provided that the block sizes are small enough, say 32$\times$32 or
64$\times$64 for todays' typical processor and memory architectures,
both the input and output blocks will reside in the high-speed cache
memory and fast read/write access to the cache will dramatically
improve performance.  Figure \ref{fig:blocking} shows a schematic
diagram of how the product $M=Y_1^\dag Y_2$ would be carried out in a
blocked manner.

\begin{figure}[t!b!]
\begin{picture}(6.5,3)(0,0)
\put(0,0){\framebox(6.5,3){}}
\put(0.5,1.5){\oval(0.2,1)[l]}
\put(1.5,1.5){\oval(0.2,1)[r]}
\put(1.0,1.0){\line(0,1){1}}
\put(0.5,1.5){\line(1,0){1}}
\put(2.0,1.5){\makebox(0,0){\large =}}
\put(2.0,2.8){\makebox(0,0){\large =}}
\put(0.75,1.75){\makebox(0,0){$m_{00}$}}
\put(1.25,1.75){\makebox(0,0){$m_{01}$}}
\put(0.75,1.25){\makebox(0,0){$m_{10}$}}
\put(1.25,1.25){\makebox(0,0){$m_{11}$}}
\put(1.0,2.8){\makebox(0,0){\large $M$}}
\put(3.5,2.0){\oval(2,0.2)[t]}
\put(3.5,1.0){\oval(2,0.2)[b]}
\put(2.5,1.5){\line(1,0){2}}
\put(4.0,1.0){\line(0,1){1}}
\put(3.5,1.0){\line(0,1){1}}
\put(3.0,1.0){\line(0,1){1}}
\put(2.75,1.75){\makebox(0,0){$a_{00}$}}
\put(3.25,1.75){\makebox(0,0){$a_{01}$}}
\put(3.75,1.75){\makebox(0,0){$a_{02}$}}
\put(4.25,1.75){\makebox(0,0){$a_{03}$}}
\put(2.75,1.25){\makebox(0,0){$a_{10}$}}
\put(3.25,1.25){\makebox(0,0){$a_{11}$}}
\put(3.75,1.25){\makebox(0,0){$a_{12}$}}
\put(4.25,1.25){\makebox(0,0){$a_{13}$}}
\put(3.5,2.8){\makebox(0,0){\large $Y_1^\dag$}}
\put(4.75,1.5){\makebox(0,0){\large $\times$}}
\put(5,1.5){\oval(0.2,2)[l]}
\put(6,1.5){\oval(0.2,2)[r]}
\put(5.5,0.5){\line(0,1){2}}
\put(5.0,2.0){\line(1,0){1}}
\put(5.0,1.5){\line(1,0){1}}
\put(5.0,1.0){\line(1,0){1}}
\put(5.25,2.25){\makebox(0,0){$b_{00}$}}
\put(5.75,2.25){\makebox(0,0){$b_{01}$}}
\put(5.25,1.75){\makebox(0,0){$b_{10}$}}
\put(5.75,1.75){\makebox(0,0){$b_{11}$}}
\put(5.25,1.25){\makebox(0,0){$b_{20}$}}
\put(5.75,1.25){\makebox(0,0){$b_{21}$}}
\put(5.25,0.75){\makebox(0,0){$b_{30}$}}
\put(5.75,0.75){\makebox(0,0){$b_{31}$}}
\put(5.5,2.8){\makebox(0,0){\large $Y_2$}}
\put(3.5,0.4){\makebox(0,0){\large\parbox{3in}{\[m_{ij} = \sum_k
a_{ik}b_{kj}\]}}}
\end{picture}
\caption[Blocked matrix multiplication]{\label{fig:blocking}Blocked
matrix multiplication -- A schematic of the blocked matrix
multiplication method applied to computing the product $M=Y_1^\dag
Y_2$.  The blocks $m_{ij}$, $a_{ij}$, and $b_{ij}$ are matrices of
small size (e.g. $32\times 32$).  Our schematic shows how each of the
matrices is partitioned into blocks and how the result blocks $m_{ij}$
are to be computed.}
\end{figure}

Please note that due to blocking, the task of optimization is now also
modularized.  We need only write a single, low-level, block-block
matrix multiplication routine that should be highly optimized.  All
higher-level matrix multiplication functions will then loop over
blocks of the input and output data and copy the contents to small,
in-cache matrices which are then multiplied by calling the low-level
multiplier.

Applying these ideas to our $M=Y_1^\dag Y_2$ example, the rewritten
code for the\\ {\tt column\_bundle\^{}column\_bundle} operator takes
the following form:

{\tt\begin{tabbing}
\codeindent\=friend matrix 
operator\^{}(column\_bundle \&Y\_1, column\_bundle \&Y\_2)\\
\>\{\ \ \=\\
\>\>matrix M(Y\_1.n\_columns,Y\_2.n\_columns);  // M=Y\_1\^{}Y\_2
holds the result\\
\\
\>\>int B = 32; // Pick a reasonable block size\\
\\
\>\>matrix in1(B,B),in2(B,B),out(B,B); // input and output blocks\\
\\
\>\>int ib,jb,kb,i,j,k;\\
\\
\>\>// loop over blocks of the output\\
\>\>for\= (ib=0; ib < Y\_1.n\_columns; ib+=B)\\
\>\>\>for\= (jb=0; jb < Y\_2.n\_columns; jb+=B) \{\\
\\
\>\>\>\>// zero the output block\\
\>\>\>\>for\= (i=0; i < B; i++)\\
\>\>\>\>\>for\= (j=0; j < B; j++)\\
\>\>\>\>\>\>out(i,j) = 0;\\
\\
\>\>\>\>// loop over blocks to be multiplied\\
\>\>\>\>for\= (kb=0; kb < Y\_1.nrows; kb+=B) \{\\
\\
\>\>\>\>\>// load data into input blocks\\ 
\>\>\>\>\>for\= (i=0; i < B; i++)\\ 
\>\>\>\>\>\>for\= (k=0; k < B; k++) \{\\ 
\>\>\>\>\>\>\>in1(i,k) = conjugate(Y\_1(kb+k,ib+i));\\
\>\>\>\>\>\>\>in2(i,k) = Y\_2(kb+k,jb+i);\\ 
\>\>\>\>\>\}\\ 
\\
\>\>\>\>\>// perform the block multiplication\\
\>\>\>\>\>low\_level\_mutliply(in1, in2, out, B);\\ 
\>\>\>\>\}\\ 
\\
\>\>\>\>// write the result to memory\\ 
\>\>\>\>for\= (i=0; i < B; i++)\\ 
\>\>\>\>\>for\= (j=0; j < B; j++)\\ 
\>\>\>\>\>\>M(ib+i,jb+j) = out(i,j);\\ 
\>\>\}\\
\\
\>\>return M;\\
\>\}
\end{tabbing}}

The function \texttt{low\_level\_multiply} performs the block-block
multiplication of the input blocks and accumulates into the output
block.  Not only the \texttt{column\_bundle\^{}column\_bundle}
operator but {\em all} matrix multiplications can be blocked in a
similar way and will thus call the low-level multiplier.

\begin{figure}[t!b!]
\begin{center}
\resizebox{4.5in}{!}{\includegraphics{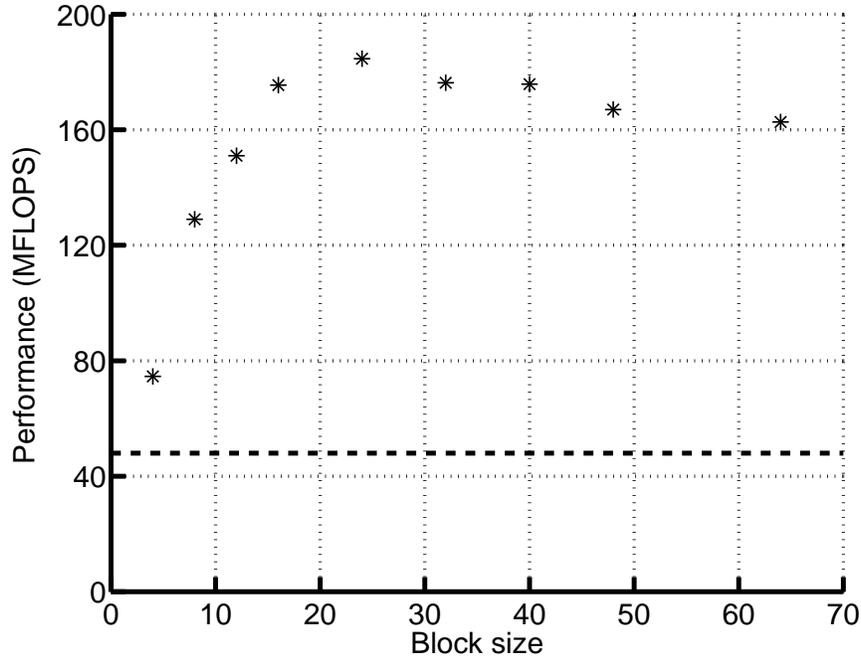}}
\caption[Matrix-multiplication FLOP rates (single
processor)]{\label{fig:blockingmFLOPS}Matrix-multiplication FLOP rates
(single processor) -- Effect of blocking on computational performance
for the matrix product $M=Y_1^\dag Y_2$, where $Y_1$ and $Y_2$ are
10,000$\times$200 complex-valued matrices.  The horizontal axis shows
the size of the square blocks used for the block-multiplication scheme
described in the text.  The vertical axis is the performance in mega
floating-point operations per second (MFLOPS).  The horizontal dashed
line shows the rate for a non-blocked ``naive'' multiplication routine
(see text).}
\end{center}
\end{figure}

The simplest implementation for \texttt{low\_level\_multiply} is the
straightforward one:

{\tt\begin{tabbing}
\codeindent\=void low\_level\_multiply(\=matrix \&in1, matrix \&in2,\\
\>\> matrix \&out, int B)\\
\>\{\ \ \=\\
\>\>int i,j,k;\\
\>\>for\= (i=0; i < B; i++)\\
\>\>\>for\= (j=0; j < B; j++) \{\\
\>\>\>\>complex z = 0;\\
\>\>\>\>for\= (k=0; k < B; k++)\\
\>\>\>\>\>z += in1(i,k)*in2(j,k);\\
\>\>\>\>out(i,j) += z;\\
\>\>\>\}\\
\>\}
\end{tabbing}}

The use of this simple low-level multiplier combined with blocking can
provide a tremendous improvement.  Figure \ref{fig:blockingmFLOPS}
shows a plot of the performance of the $M=Y_1^\dag Y_2$
blocked-multiplication as a function of the block size when run on a
333 MHz SUN Ultrasparc 5 microprocessor.  For comparison, the
performance of the ``naive'' code with no blocking, which was
presented above, is also indicated in the figure.  Initially, the
performance increases dramatically with increasing block size due to
more effective caching.  However, there is an optimal size above which
performance decreases because the blocks become too large to fit
effectively into the cache.  On most computational platforms that we
have had experience with, this simple blocked multiplier runs at at
least half the peak theoretical rate of the processor, as exemplified
by the results in the figure.  Further speedups can be found by
rewriting the \texttt{low\_level\_multiply} routine so as to use
register variables (as we have done and found up to 40\% performance
enhancements) or by using a hierarchical access pattern to memory
which can provide better performance if the memory system has multiple
levels of caching.

\subsection{Parallelization}
\label{sec:parallel}

Once the computer code has been optimized to perform well on a single
processor, the computation can be divided among multiple processors.
We now discuss how this division can be achieved effectively.

The architectures of modern parallel supercomputers can generally be
divided into two categories: shared-memory (SMP) versus
distributed-memory (DMP) processors.  In the SMP case, a set of
identical processors share access to a very large repository of
memory.  The main advantages of shared memory are that the processors
can access whatever data they need, and that, with judicious choice of
algorithms, very little inter-processor communication is required.  In
addition, only small changes are required to parallelize a serial
code: the computational task is divided among the processors, and each
processor executes the original serial code on the portion of the data
allotted to it.  However, as the number of processors increases, the
traffic for accessing the main memory banks increases dramatically and
the performance will stop to scale well with the number of processors
utilized.  Nevertheless, many mid-sized problems are well suited to
SMPs and can take full advantage of the key features of SMP systems.
Examples of such calculations can be found in \cite{Sibars,Sidisloc}.

The largest of today's supercomputers have distributed memory: each
processor has a private memory bank of moderate size to which it has
exclusive access, and the processors communicate with each other by
some message passing mechanism.  The main advantages of DMP are
scalability and heterogeneity, as the processors need not all be
identical nor located in close physical proximity.  However, the price
paid is the necessity of an inter-processor communication mechanism
and protocol.  In addition, computer algorithms may have to be
redesigned in order to minimize the required communication.
Furthermore, a slow communication network between processors can
adversely affect the overall performance.

We will describe, in outline, the strategies we employ for both SMP
and DMP architectures, and we will continue to examine the case of
the\\ {\tt column\_bundle\^{}column\_bundle} matrix multiplication as
a specific example.  As our results for actual calculations will
demonstrate, we only need to parallelize two main operations in the
entire code, (a) the application of basis-dependent operators such as
$\I$ or $L$ to \texttt{column\_bundle}s (which is trivial) and (b) the
matrix products involving \texttt{column\_bundle}s such as $Y_1^\dag
Y_2$ or $YU^{-1/2}$, to obtain excellent or highly satisfactory
performance on SMP and DMP systems.

\subsubsection{Shared-memory (SMP) architectures}
\label{sec:SMPparallel}

Since all processors in an SMP computer have access to all the data in
the computation, the parallelization of the basis-dependent operators
is trivial.  For example, the operation $\I C$ involves the
application of $\I$ to each column of $C$ separately, and the columns
can be divided equally among the processors.  This leads to near
perfect parallelization as none of the processors read from or write
to the same memory locations.

Based on the discussion of Section~\ref{sec:FLOPcount}, for large
problems, the most significant gains for parallelization involve the
basis-independent matrix-products.  Below, we focus on the {\tt
column\_bundle\^{}column\_bundle} operation as a case study.  For this
operation, it is impossible to distribute the task so that all
processors always work on different memory segments.  However, we
divide up the work so that no two processors ever write to the same
memory location: not only does this choice avoid possible errors due
to the synchronizations of simultaneous memory writes, it also avoids
performance degradation due to cache-flushing when memory is written
to.

Consider the matrix product $M = Y_1^\dag Y_2$, which we have
implemented as a blocked matrix multiplication.  Parallelization is
achieved simply by assigning each processor to compute a subset of the
output blocks.  The main program spawns a set of subordinate tasks
(termed threads) whose sole purpose is to compute their assigned
output blocks and to write the results to memory.  The main program
waits for all threads to terminate before continuing onwards.
Referring to Figure \ref{fig:blocking}, this strategy corresponds to
distributing the computation of the blocks $m_{ij}$ among the
processors, and since memory is shared, all processors have access to
all of the data describing $Y_1$ and $Y_2$ at all times.  For large
problem sizes, the overhead in spawning and terminating the threads is
far smaller than the work needed to compute the results, so the
simplicity of this strategy does not sacrifice performance.

We now present the code which accomplishes this parallelized matrix
product in order to highlight the ease with which such parallelizations
can be performed.  In the interest of brevity, we assume that the
number of columns of $Y_1$ is a multiple of the number of threads
launched.  Parallelization is affected by assigning different threads
to compute different rows of the result $M=Y_1^\dag Y_2$.

{\tt\begin{tabbing}
\codeindent\=friend matrix operator\^{}(column\_bundle \&Y\_1,
column\_bundle \&Y\_2)\\
\>\{\ \ \ \=\\
\>\>matrix M(Y\_1.n\_columns,Y\_2.n\_columns);  // M=Y\_1\^{}Y\_2 holds
the result\\
\\
\>\>int n\_threads = 8;  // a reasonable number of threads\\
\\
\>\>int thread\_id[n\_threads];\\
\>\>int i, start, n;\\
\\
\>\>for \=(i=0; i < n\_threads; i++) \{\\
\\
\>\>\>// The set of rows of M that this thread should compute\\
\>\>\>n = Y\_1.n\_columns/n\_threads;\\
\>\>\>start = i*n;\\
\\
\>\>\>// Launch a thread that runs the routine calc\_rows\_of\_M\\
\>\>\>// and pass it the remaining arguments.  Store the thread ID.\\
\>\>\>thread\_id[i] = launch\_thread(\=calc\_rows\_of\_M, Y\_1, Y\_2, M,\\
\>\>\>\>start, n);\\
\>\>\}\\
\\
\>\>// Wait for the threads to terminate\\
\>\>for \=(i=0; i < n\_threads; i++)\\
\>\>\>wait\_for\_thread(thread\_id[i]);\\
\\
\>\>return M;\\
\>\}
\end{tabbing}}

The routine \texttt{calc\_rows\_of\_M(Y\_1,Y\_2,M,start,n)} computes
rows \texttt{start} through\\ \texttt{start+n-1} of \texttt{M}, where
\texttt{M=Y\_1\^{}Y\_2}.  The routine's algorithm is identical to that
of the blocked multiplier presented in the previous section.  The only
change required is to have the \texttt{ib} loop index start at
\texttt{start} and end at \texttt{start+n-1}.

\begin{figure}[t!b!]
\begin{center}
\resizebox{4.5in}{!}{\includegraphics{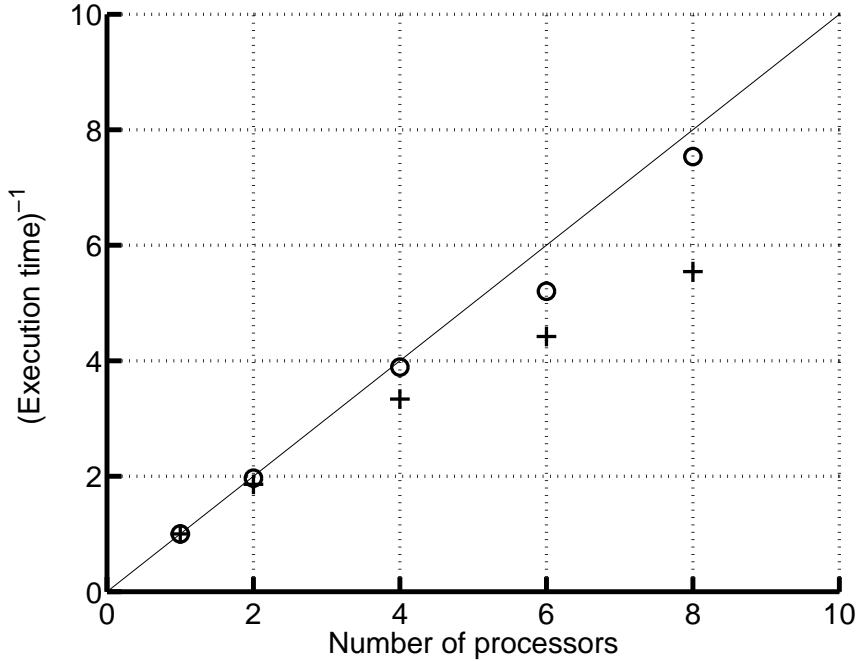}}
\caption[Scaling for SMP parallelization]{\label{fig:smp1}Scaling for
SMP parallelization -- Performance of our SMP parallelized plane-wave
code for a 128 atom silicon system.  We show the performance of the
parallelized $M=Y_1^\dag Y_2$ multiplications (circles) and the overall
code (pluses) as a function of the number of processors .  In both
cases, performance has been normalized to the respective
performance with a single processor.  The straight line with slope of
unity represents ideal scaling for perfect parallelization.}
\end{center}
\end{figure}

We parallelize other matrix multiplications involving
\texttt{column\_bundle}s analogously.  In addition, we parallelize the
application of the basis-dependent operators as discussed above.  For
a realistic study of performance and scaling, we consider a system of
bulk silicon with 128 atoms in the unit cell.  We use a plane-wave
cutoff of 6 Hartrees (12 Ryd) which leads to a basis of size
$N=11797$.  We use Kleinmann-Bylander \cite{KB} non-local
pseudopotentials with $p$ and $d$ non-local projectors to eliminate
the core states, and we have $n=256$ bands of valence electrons.  We
sample the Brillouin zone at $k=0$.  In Figure \ref{fig:smp1}, we show
a plot of the performance of the parallelized $M=Y_1^\dag Y_2$
multiplication as well as the overall performance of the code for a
single conjugate-gradient step as a function of the number of
processors employed.  The calculation was run on a Sun Ultra HPC 5000
with eight 167 MHz Ultrasparc 4 microprocessors and 512MB of memory.

As the figure shows, the parallelized $M=Y_1^\dag Y_2$ matrix
multiplication shows near ideal scaling.  There are a number of
reasons for this behavior: (1) since different processors always write
to different segments of memory, the algorithm does not suffer from
memory write-contentions, (2) the inputs $Y_1$ and $Y_2$ are never
modified so that memory reads are cached effectively, and (3) the
problem size is large enough so that each processor's workload is much
larger than the overhead required to distribute the work among the
processors.  This scaling is all the more impressive because when
using eight processors, each processor performs the multiplications at
140 MFLOPS or at 83\% of the processor clock rate.

\begin{figure}[t!b!]
\begin{center}
\resizebox{4.5in}{!}{\includegraphics{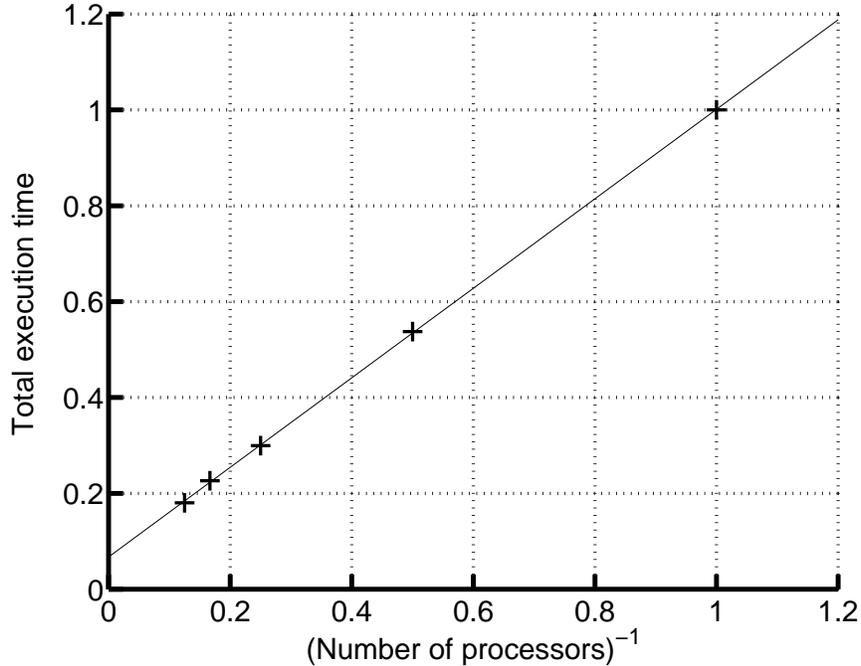}}
\caption[Amdahl's analysis of SMP scaling]{\label{fig:smp2}Amdahl's
analysis of SMP scaling -- Total execution time of the SMP parallelized
plane-wave code for a 128 silicon atom system as a function of the
reciprocal of the number of processors used.  The line of
least-squares fit to the data points is also shown.  Execution times
are normalized in units of the execution time for a single processor.}
\end{center}
\end{figure}

The figure also displays the total performance of the code, which
shows evidence of saturation.  To understand this behavior in more
detail, we model the overall execution time in accordance with
Amdahl's law.  We assume that there exists an intrinsic serial
computation time $T_0$ which must be spent regardless of the number of
processors available.  In addition, there is an analogous parallel
computation time $\tilde T$ which, however, is divided equally among
all the processors.  Thus the total execution time will be given by
$T=T_0 + \tilde T/p$, where $p$ is the number of processors.  We show
a plot of the total execution time versus $1/p$ in Figure
\ref{fig:smp2}, and the model shows an excellent fit to the available
data.  The vertical asymptote of the least-squares fit straight line
is the extrapolated value of $T_0$, in this case approximately 10\% of
the single processor or serial execution time.  Thus the few
operations that we have chosen to parallelize do in fact constitute
the largest share of the computational burden and our parallelization
strategy is quite effective.

When we reach the data point with eight processors, the total
execution time is already within a factor of two of $T_0$, so that the
total serial and parallel components have become comparable.  Indeed,
timing various portions of the code confirms this hypothesis: for
example, with eight processors, the time needed to perform a parallel
multiplication $M=Y_1^\dag Y_2$ is equal to the time needed by the Sun
high-performance LAPACK library to diagonalize the overlap matrix $U$
(cf. Eq.~(\ref{eq:Udef})).  With eight processors, the code has each
processor {\em sustaining} an average of 134 MFLOPS or 80\% of the
processor clock rate.  We are quite satisfied with this level of
performance, but if more improvements are desired, the remaining
serial portions of the code can be further optimized and parallelized.

\subsubsection{Distributed-memory (DMP) architectures}
\label{sec:DMPparallel}

Parallelization on DMP architectures requires careful thought
regarding how the memory distribution and inter-process communications
are to be implemented.  The most memory-consuming computational
objects are the \texttt{column\_bundle}s, and for a large system, no
single processor in a DMP computer can store the entire data structure
in its local memory banks.  Therefore, we parallelize the storage of
\texttt{column\_bundle}s by distributing equal numbers of the columns
comprising a \texttt{column\_bundle} among the processors.

Given this distribution by columns, the application of basis-dependent
operators is unchanged from how it is done in a serial context: each
processor applies the operator to the columns that are assigned to it,
and perfect parallelization is achieved as no inter-processor
communication is required.  The large basis-independent matrix
products, however, are more challenging to parallelize as they
necessarily involve inter-processor communication.

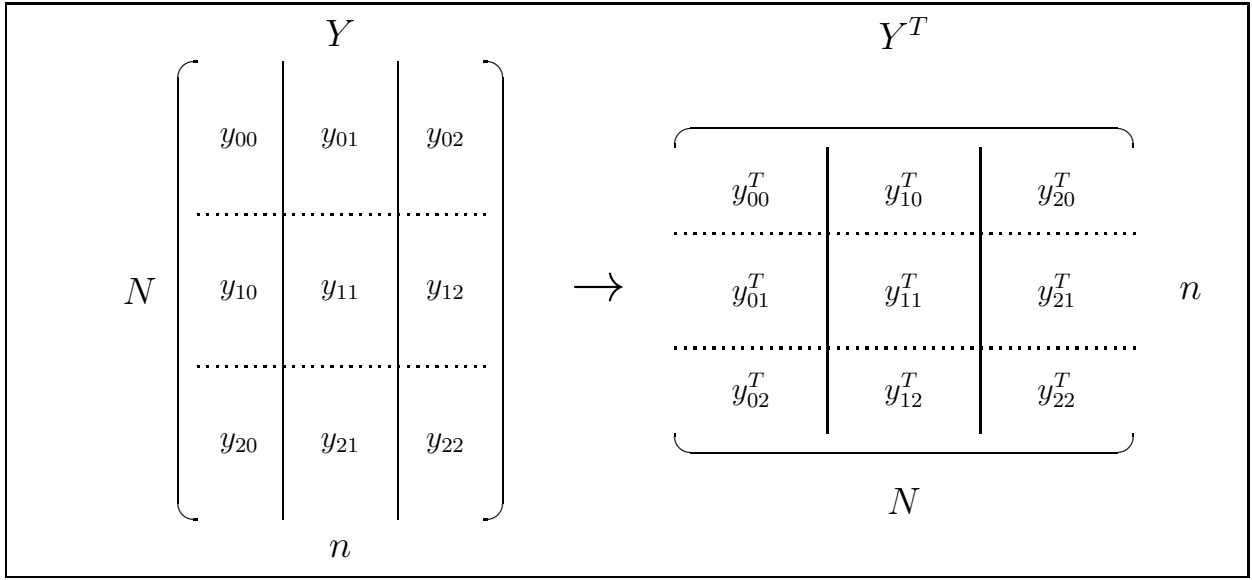
\begin{figure}[t!b!]
\begin{picture}(6.5,3)(0,0)
\put(0,0){\framebox(6.5,3){}}
\put(1.0,1.5){\oval(0.2,2.4)[l]}
\put(2.5,1.5){\oval(0.2,2.4)[r]}
\put(1.45,0.3){\line(0,1){2.4}}
\put(2.05,0.3){\line(0,1){2.4}}
{\thicklines \qbezier[30](1.0,1.1)(1.75,1.1)(2.5,1.1)}
{\thicklines \qbezier[30](1.0,1.9)(1.75,1.9)(2.5,1.9)}
\put(1.22,2.3){\makebox(0,0){$y_{00}$}}
\put(1.75,2.3){\makebox(0,0){$y_{01}$}}
\put(2.30,2.3){\makebox(0,0){$y_{02}$}}
\put(1.22,1.5){\makebox(0,0){$y_{10}$}}
\put(1.75,1.5){\makebox(0,0){$y_{11}$}}
\put(2.30,1.5){\makebox(0,0){$y_{12}$}}
\put(1.22,0.7){\makebox(0,0){$y_{20}$}}
\put(1.75,0.7){\makebox(0,0){$y_{21}$}}
\put(2.30,0.7){\makebox(0,0){$y_{22}$}}
\put(1.75,2.85){\makebox(0,0){\large $Y$}}
\put(1.75,0.15){\makebox(0,0){\large $n$}}
\put(0.70,1.5){\makebox(0,0){\large $N$}}
\put(3.1,1.5){\makebox(0,0){\LARGE $\rightarrow$}}
\put(4.7,0.75){\oval(2.4,0.2)[b]}
\put(4.7,2.25){\oval(2.4,0.2)[t]}
\put(4.3,0.75){\line(0,1){1.5}}
\put(5.1,0.75){\line(0,1){1.5}}
{\thicklines \qbezier[48](3.5,1.2)(4.7,1.2)(5.9,1.2)}
{\thicklines \qbezier[48](3.5,1.8)(4.7,1.8)(5.9,1.8)}
\put(3.9,2.03){\makebox(0,0){$y_{00}^T$}}
\put(4.7,2.03){\makebox(0,0){$y_{10}^T$}}
\put(5.5,2.03){\makebox(0,0){$y_{20}^T$}}
\put(3.9,1.50){\makebox(0,0){$y_{01}^T$}}
\put(4.7,1.50){\makebox(0,0){$y_{11}^T$}}
\put(5.5,1.50){\makebox(0,0){$y_{21}^T$}}
\put(3.9,0.98){\makebox(0,0){$y_{02}^T$}}
\put(4.7,0.98){\makebox(0,0){$y_{12}^T$}}
\put(5.5,0.98){\makebox(0,0){$y_{22}^T$}}
\put(4.7,2.85){\makebox(0,0){\large $Y^T$}}
\put(6.20,1.5){\makebox(0,0){\large $n$}}
\put(4.70,0.40){\makebox(0,0){\large $N$}}
\end{picture}
\caption[Transposition of distributed
matrices]{\label{fig:transpose}Transposition of distributed matrices
-- Schematic diagram showing how the transpose $Y^T$ of the
distributed \texttt{column\_bundle} $Y$ is obtained in a DMP
calculation, which in this example has $p=3$ processors.  Across all
the processors, $Y$ is $N\times n$ and $Y^T$ is $n\times N$.  Solid
vertical lines show the distribution of the columns of $Y$ or $Y^T$
among the processors, so that each processor stores a $N\times (n/p)$
block of $Y$ and a $n\times (N/p)$ block of $Y^T$.  The horizontal
dotted lines show the division of $Y$ and $Y^T$ into blocks that must
be communicated between processors to achieve the transposition: the
block $y_{ij}$ is sent from processor $j$ to processor $i$.  Processor
$i$ then transposes the block and stores it in the $ji$th section of
$Y^T$.}
\end{figure}

Consider again the product $M = Y_1^\dag Y_2$, which in terms of
components is given by
\[
M_{ij} = \sum_k (Y_1)_{ki}^* (Y_2)_{kj}\,.
\]
The column-wise parallel distribution of $Y_1$ and $Y_2$ means that
the full range of the $i$ and $j$ indices is distributed among the
processors while the full range of the $k$ index is accessible locally
by each processor.  Since $Y_1$ and $Y_2$ are $N\times n$, each
processor has a $N\times (n/p)$ block (i.e. $n/p$ columns of length
$N$) in its local memory, where $p$ is the number of processors.
Unfortunately, computing $M$ using this column-wise distribution
requires a great deal of inter-processor communication.  For example,
the processor computing the $i$th row of $M$ requires knowledge of
{\em all} the columns of $Y_2$, so that in total, the $Nn$ data items
describing $Y_2$ will have to be sent $p-1$ times between all the
processors.

\begin{figure}[t!b!]
\begin{center}
\resizebox{4.5in}{!}{\includegraphics{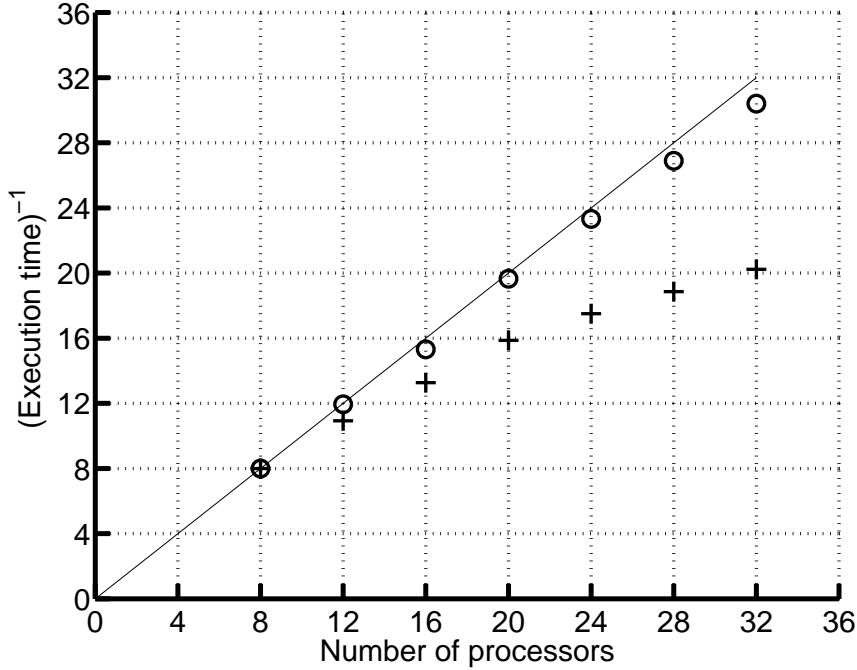}}
\caption[Scaling of DMP parallelization]{\label{fig:dmp1}Scaling of
DMP parallelization -- Performance of the DMP parallelized plane-wave
code for a 256 silicon atom system as a function of the number of
processors for the parallelized $M=Y_1^\dag Y_2$ multiplication
(circles) and for the code overall (pluses).  In both cases, the
performance has been normalized to the respective performance with a
single processor based on extrapolation from the eight processor run.
The straight line with slope of unity represents ideal scaling for
perfect parallelization.}
\end{center}
\end{figure}

A more efficient communication pattern can be developed that avoids
this redundancy.  Denoting the transpose of $Y$ by $Y^T$, the matrix
product can be rewritten as
\[
M_{ij} = \sum_k (Y_1^T)_{ik}^* (Y_2^T)_{jk}\,.
\]
Hence, if we first transpose $Y_1$ and $Y_2$, then the column-wise
distribution of the transposed \texttt{column\_bundle}s insures that
the full range of the $i$ and $j$ indices can be accessed locally on
each processor while the full range of the $k$ index is now
distributed among the processors.  Figure \ref{fig:transpose} presents
a schematic of how the transposition of a \texttt{column\_bundle} $Y$
is be accomplished: each processor has a $N\times (n/p)$ block of $Y$
whose contents it sends to $p-1$ other processors as $p-1$ blocks of
size $(N/p)\times (n/p)$.  Next, each processor receives $p-1$ similar
blocks sent to it by other processors, transposes them, and stores
them in the appropriate sections of $Y^T$.  Each processor sends or
receives only $Nn(p-1)/p^2$ data items, and a total of $Nn(p-1)/p$
data items are communicated among all the processors.

The computation of $M$ in the transposed mode has the same operation
count as in the non-transpose mode (i.e. $O(n^2N)$ operations), which
when distributed across processors, amounts to $O(n^2N/p)$ operations
per processor.  Of course, we block the matrix multiplications on each
processor to ensure the best performance.  Finally, a global sum
across all the processors' $n\times n$ resulting matrices is required
to obtain the final answer $M$, and this requires $\log_2 p$
communications of size $n^2$ when using a binary tree.

Assuming that processors can simultaneously send and receive data
across the network, the time required to perform the transpose is
$O(nN/p)$, the time needed to perform the multiplications is
$O(n^2N/p)$, and the time required to perform the final global sum is
$O(n^2\log_2 p)$.  For large problem sizes, the time needed to perform
the multiplications will always be larger than the time required for
the communications by a factor of $\sim n$.  Thus interprocessor
communications are, in the end, never an issue for a sufficiently
large physical system, and the computation will be perfectly
parallelized in the asymptotic limit of an infinitely large system.

\begin{figure}[t!b!]
\begin{center}
\resizebox{4.5in}{!}{\includegraphics{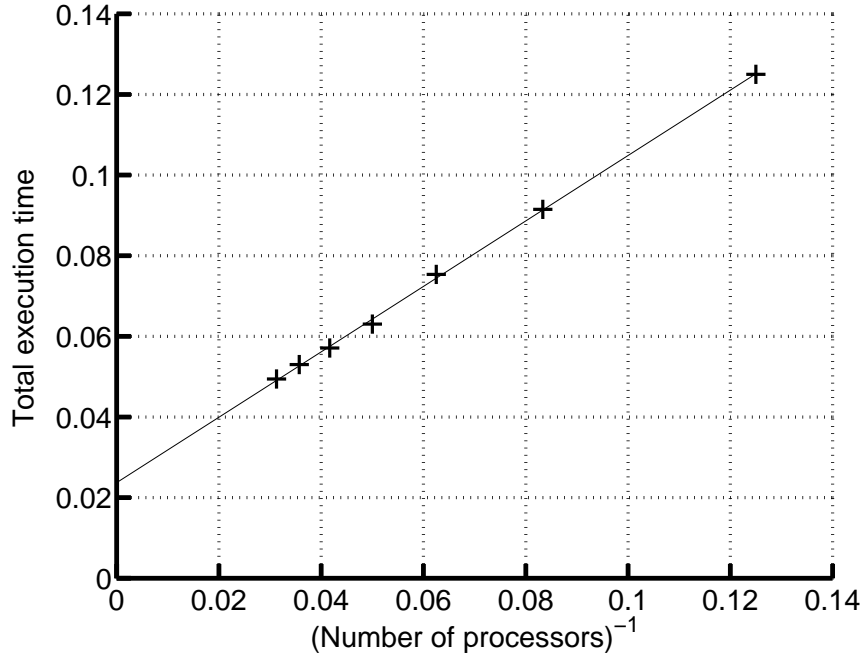}}
\caption[Amdahl's analysis of DMP scaling]{\label{fig:dmp2}Amdahl's
analysis of DMP scaling -- Total execution time of the DMP
parallelized plane wave code for a 256 silicon atom system as a
function of the reciprocal of the number of processors utilized.  The
line of least-squares fit to the data points is shown as well.
Execution times are normalized in units of the execution time for a
single processor as extrapolated based on the eight processor run.}
\end{center}
\end{figure}

Figure~\ref{fig:dmp1} shows a plot of the performance of our DMP
parallelized plane-wave code for a single conjugate-gradient step when
run on a 256 atom silicon cell.  The choice of pseudopotential,
cutoff, and $k$-point sampling are the same as for the SMP calculation
above.  The basis set is of size $N=23563$ and the system has $n=512$
valence bands.  The calculations were run on an IBM SP2 system with
336 nodes, and each node has four 332 MHz Power2 Architecture RISC
System/6000 processors and 1.536 GB of memory.  Again, we see
excellent and near perfect scaling for the parallelized matrix
multiplications, validating our claim that the transposition approach
combined with the large system size provides for very good
parallelization.  With 32 processors, each parallelized multiplication
runs at an average rate of 188 MFLOPS per processor (57\% of the
processor clock rate), which is impressive given the fact that the
processors are busily communicating the data required for the
transpositions and global sums.

The plot also shows the saturation of the total performance of the
code with increasing number of processors.  With eight processors, the
overall performance translates into an average rate of 254 MFLOPS per
processor (76\% of the clock rate), whereas with 32 processors the
rate has reduced to 160 MFLOPS per processor (48\% of the clock rate).
Clearly, the serial portions of the calculation begin to contribute
significantly to the run time of the code.  Following the discussion
of the SMP results above, Figure~\ref{fig:dmp2} presents a plot of the
total execution time versus inverse number of processors.  The
extrapolated serial time $T_0$ in this case is only 2-3\% of the total
theoretical run time on a single processor, which shows how
effectively the calculation has been parallelized.  In addition, for
the 32 processor run, the total run time is only two times larger that
$T_0$, signalling the end of significant gains from the use of more
processors.

\section{Acknowledgements}

We would like to thank Jason A. Cline for his work in implementing the
LSDA functional.  We are immensely grateful to Tairan Wang for his
work in creating the MPI implementation of the software and to Kenneth
P. Essler for creating the MPI distributed transposition routine and
the accompanying efficient matrix multiplication routines.

This work was supported primarily by the MRSEC Program of the National
Science Foundation under award number DMR 94-00334, by the Alfred
P. Sloan Foundation (BR-3456), and also by an ASCI ASAP Level 2 grant
(contract numbers \#B338297 and \#B347887). Code development was
carried out on the MIT Xolas prototype SMP cluster as well as on the
MIT Pleiades Alpha cluster.  The calculations were carried out on the
Xolas cluster and on the ASCI Blue Pacific Teraflops IBM SP2 platform.
This work made use of the Cornell Center for Materials Research Shared
Experimental Facilities, supported through the NSF MRSEC program
(DMR-9632275).

%
%
\newpage

\appendix
\section{Plane-wave implementation of the basis-dependent operators}
\label{appendix:implementpw}

A widely used basis-set for \abi calculation has been the plane-wave
basis set \cite{RMP}.  Plane waves are ideally suited for periodic
calculations that model the bulk of a crystalline material.  In
addition, plane waves provide uniform spatial resolution throughout
the entire simulation cell, and the results of the calculations can be
converged easily by simply increasing the number of plane waves in the
basis set.  We will use plane waves as a concrete example of how the
basis-dependent operators of Section~\ref{sec:basisdepops} are to be
implemented.

Given the lattice vectors of a periodic supercell, we compute the
reciprocal lattice vectors and denote the points of the reciprocal
lattice by the vectors $\{G\}$.  Each element of our basis set
$\{b_\alpha(r)\}$ will be a plane wave with vector $G_\alpha$,
\[
b_\alpha(r) = { e^{iG_\alpha\cdot r} \over \sqrt{\Omega} }\,,
\]
where $\Omega$ is the real-space volume of the periodic cell.  This
basis is orthonormal, and the overlap operator $\calO$ is the identity
operator,
\[
\calO = I\,.
\]
The integrals of the basis functions $s$ are given
by
\[
s_\alpha = \sqrt{\Omega}\,\delta_{G_\alpha,0}\,,
\]
so that the $\Obar$ operator is given by
\[
\Obar_{\alpha \beta} = \delta_{\alpha,\beta} -
\delta_{G_\alpha,0}\cdot\delta_{G_\beta,0}\,.
\]
The Laplacian operator $L$ is diagonal in this basis and is given by
\[
L_{\alpha \beta} = -\|G_\alpha\|^2\,\delta_{\alpha\beta}\,.
\]

The forward transform $\I$ is given by a Fourier transformation.
Specifically, for a point $p$ on the real-space grid, we have that
\[
\I_{p \alpha} = {e^{iG_\alpha\cdot p} \over \sqrt{\Omega} }\,.
\]
Consider applying $\I$ to the column vector $\chi$ and evaluating the
result at the point $p$:
\[
(\I\chi)_p = \sum_\alpha \I_{p \alpha}\,\chi_\alpha =
{1\over \sqrt{\Omega}}\sum_\alpha e^{iG_\alpha\cdot p}\,\chi_\alpha
\,.
\]
This is the forward Fourier transform of $\chi$.

For the case of plane waves, the inverse transform $\J$ can be chosen
to be the inverse of $\I$, $\J =\I^{-1}$, as per the discussion of
Section~\ref{sec:basisdepops}.  It follows that
\[
\J  = \I^{-1} = \left({\Omega \over N}\right)\Idag
\ \ \ \mbox{or}\ \ \ \ \J _{\alpha p} = {\sqrt{\Omega}
\over N}e^{-iG_\alpha\cdot p}\,,
\]
where $N$ is the number of points in the real-space grid. Applying
$\J$ to a column vector $\xi$, we have
\[
(\J\xi)_{\alpha} = {\sqrt{\Omega} \over N}\sum_p
e^{-iG_\alpha\cdot p}\,\xi_p\,.
\]
Thus $\J$ is a reverse Fourier transform.  The operators $\Idag$ and
$\Jdag$ are also Fourier transforms with appropriate scaling factors.
Computationally, Fast Fourier Transforms \cite{fftw,fft} can be used
to implement these operators most efficiently.

The last remaining basis-dependent item is the ionic potential
$V_{ion}$.  For periodic systems, this potential is a periodic sum of
atomic potentials $V_{at}(r)$,
\[
V_{ion}(r) = \sum_{R,I} V_{at}(r-R-r_I)\,,
\]
where $R$ ranges over the real-space lattice sites, $I$ ranges over
the atoms in the unit cell, and $r_I$ is the position of the $I$th
atom.  Based on this, the elements of the vector $V_{ion}$ of
Eq.~(\ref{eq:Viondef}) are given by
\[
(V_{ion})_\alpha = {S(G_\alpha)\hat{V}_{at}(G_\alpha) \over
\sqrt{\Omega}}\,,
\]
where the structure factor $S(q)$ is given by
\[
S(q) \equiv \sum_I e^{-iq\cdot r_I}\,,
\]
and the Fourier transform of the atomic potential $\hat{V}_{at}$ is
defined by
\[
\hat{V}_{at}(q) \equiv \int d^3r\ e^{-iq\cdot r}\, V_{at}(r)\,.
\]

\section{Non-local potentials}
\label{appendix:nlpots}

In this section we show how non-local potentials can easily be
incorporated into the DFT++ formalism.  The total non-local potential
operator $\hat{V}$ for a system is given by a sum over each atom's
potential,
\[
\hat{V} = \sum_I \hat{V}_I\, ,
\]
where $\hat{V}_I$ is the non-local potential of the $I$th atom.  The
non-local energy is given by the expectation of $\hat{V}$ over all the
occupied states $\psiset$ with fillings $f_i$,
\[
E_{nl} = \sum_i f_i \langle\psi_i|\hat{V}|\psi_i\rangle = 
\sum_I \sum_i f_i \langle\psi_i|\hat{V}_I|\psi_i\rangle\,.
\]
Given the linearity of $E_{nl}$ with respect to the atoms $I$, in our
discussion below we will only consider the case of a single atom and
will drop the index $I$.  The results below can then be summed over
the atoms to provide general expressions for multiple atoms.

Using the expansion coefficients $C_{\alpha i}$ of
Eq.~(\ref{eq:defCphi}), we rewrite the non-local energy as
\begin{equation}
E_{nl} = \sum_i f_i \langle\psi_i|\hat{V}|\psi_i\rangle = \sum_{i,
\alpha,\beta} f_i C^*_{\alpha i}
\langle\alpha|\hat{V}|\beta\rangle C_{\beta i} =
\mbox{Tr}(VCFC^\dag) = \mbox{Tr}(VP)\ ,
\label{eq:VPexpr}
\end{equation}
where $|\alpha\rangle$ is the ket representing the basis function
$b_\alpha(r)$, $P$ is the single-particle density matrix of
Eq.~(\ref{eq:Pdef}), the matrix $F$ was defined as
$F_{ij}=\delta_{ij}f_i$, and the matrix elements of the non-local
potential are defined as
\[
V_{\alpha \beta} \equiv \langle\alpha|\hat{V}|\beta\rangle = \int
d^3r \int d^3r'\ b^*_\alpha(r)\, V(r,r'\,)\, b_\beta(r'\,)\,.
\]
The matrix $V$ clearly depends on both the basis set and the
potential.  The contribution of the non-local potential to the total
Lagrangian of Eq.~(\ref{eq:lagrexpr2}) is given simply by
$\mbox{Tr}(VP)$.  Following the derivations of Eqs.~(\ref{eq:HdP}) and
(\ref{eq:Hspdef}), we see that the single-particle Hamiltonian $H$ is
modified only by the addition of $V$,
\begin{equation}
H = -{1\over 2}L + \Idag [\mbox{Diag }V_{sp}]\,\I + V\,.
\label{eq:HandV}
\end{equation}

We now write the potentials in separable form,
\begin{equation}
\hat{V} = \sum_{s,s'} |s\rangle M_{ss'}\langle s'|\ ,
\label{eq:Vnlseparable}
\end{equation}
where $s$ and $s'$ range over the quantum states of the atom,
$M_{ss'}$ are matrix elements specifying the details of the potential,
and $|s\rangle$ is the ket describing the contribution of the $s$th
quantum state to the potential.  Typical choices of $s$ are the
traditional atomic state labels $nlm$ and possibly the spin $\sigma$.
Once we define the matrix elements $K_{\alpha
s}\equiv\langle\alpha|s\rangle$, which are again basis-dependent, we
find two equivalent forms for $E_{nl}$,
\begin{eqnarray}
E_{nl} & = & \sum_{i,s,s'} f_i \langle\psi_i|s\rangle M_{ss'} \langle
s'|\psi_i\rangle
= \sum_{i,s,s',\alpha,\beta} f_i C^*_{\alpha i}K_{\alpha
s}M_{ss'}K^*_{\beta s'}C_{\beta i}\nonumber\\
& = & \mbox{Tr}\left[M(K^\dag C)F(K^\dag C)^\dag\right] =
\mbox{Tr}\left[KMK^\dag P\right]\ .
\label{eq:KMKdagP}
\end{eqnarray}
The first form involving $K^\dag C$ is most useful for efficient
computation of the energy, and the second form is most useful for
derivation of the gradient (cf. the discussion of
Eqs.~(\ref{eq:lagrexpr1}) and (\ref{eq:lagrexpr2})).  The energetic
contribution to the Lagrangian is given by Eq.~(\ref{eq:KMKdagP}) and
the contribution to the Hamiltonian $H$ is $KMK^\dag$, which replaces
$V$ in Eqs.~(\ref{eq:VPexpr}) and (\ref{eq:HandV}).

A further specialization involves the popular case of
Kleinmann-Bylander potentials \cite{KB} where the double sum in
Eq.~(\ref{eq:Vnlseparable}) is reduced to a single sum over $s$.
Thus, the matrix $M$ is diagonal with elements $m_s$.  The expression
for the total non-local energy, this time including the sum over atoms
$I$, is
\[
E_{nl} = \sum_{I,s} m_{Is} (K_{Is}^\dag C)F(K_{Is}^\dag C)^\dag =
\mbox{Tr}\left[ \left(\sum_{I,s} m_{Is} K_{Is}K_{Is}^\dag\right)
P\right]\,.
\]
Unfortunately, this expression is not very efficient for evaluating
the energy of a system with many atoms, as the sum on $I$ is large but
the matrix $K_{Is}^\dag C$ only has a single row.  This limits our
ability to exploit the cache effectively (which only occurs for large
matrix sizes).

We can rewrite the above energy expression so as to employ larger
matrices and thus achieve greater computational efficiency.  To do
this, we define a diagonal matrix $\bar{M}_s$ that contains the
$m_{Is}$ values for all the atoms, $(\bar{M}_s)_{IJ}\equiv
\delta_{IJ}m_{Is}$, and we define the matrices $A_s$ via
$(A_s)_{\alpha I}\equiv K_{Is,\alpha}$. We then reorganize the
previous expression for the non-local energy,
\[
E_{nl} = \sum_s \mbox{Tr}\left[\bar{M}_s (A_s^\dag C)F(A_s^\dag
C)^\dag\right] = \mbox{Tr}\left[\left(\sum_s A_s \bar{M}_s
A_s^\dag\right)P\right]\,.
\]
If we have $N$ basis functions, $n$ quantum states $\psiset$, and
$n_a$ atoms in the system, then $C$ is $N\times n$ and $A_s$ is
$N\times n_a$.  Thus, for large system sizes, the products $A_s^\dag
C$ involve matrices with large dimensions, and optimized
matrix-multiplication routines function at peak efficiency.

\section{Multiple $k$-points}
\label{appendix:kpts}

We consider the generalization of our formalism to the case of
multiple $k$-points, which arises in the study of periodic systems.
In periodic cells, the wave functions satisfy Bloch's theorem and can
be labeled by a quantum number $k$, a vector in the first Brillouin
zone.  The quantum states obey the Bloch condition
\[
\psi_k(r+R) = e^{ik\cdot R}\psi_k(r)\,,
\]
where $R$ is a lattice vector of the periodic cell.  This implies that
$\psi_k(r)=e^{ik\cdot r}u_k(r)$ where $u_k$ is a periodic function of
$r$, $u_k(r+R)=u_k(r)$.  We define the expansion coefficients $C_k$
for the vector $k$ as being those of the periodic function $u_k(r)$
and arrive at (cf. Eq.~(\ref{eq:defCphi}))
\begin{equation}
\psi_{km}(r) = e^{ik\cdot r}\sum_\alpha (C_k)_{\alpha m}\, b_\alpha(r)\,,
\label{eq:Ckdef}
\end{equation}
where the integer $m$ labels the energy bands (i.e. different states
at the same value of $k$).  The Fermi-Dirac fillings may also have a
$k$-dependence and are denoted as $f_{km}$.

In addition to $k$-vectors, calculations in periodic systems attach a
weight $w_k$ to the wave vector $k$.  The rationale is that we require
the integrals of physical functions over the Brillouin zone in order
to compute the Lagrangian, energies, and other quantities.  Ideally,
we would like to integrate a function $g(k)$ over the Brillouin zone,
but in a practical computation this must be replaced by a discrete sum
over a finite number of $k$-points with weights $w_k$.  That is, we
perform the following replacement
\[
\int d^3k\ g(k) \rightarrow \sum_k w_k\, g(k)\,.
\]

The required generalizations of the DFT++ formalism are
straightforward and are outlined below.  The density matrices
(cf. Eq.~(\ref{eq:Pdef})) now depend on $k$-points
\[
P_k = w_k C_k F_k C_k^\dag\,,
\]
where the filling matrix is $(F_k)_{mm'} =\delta_{m,m'}f_{km}$, and the
expansion coefficient matrices $C_k$ are given by
Eq.~(\ref{eq:Ckdef}).  We define the total density matrix $P$ through
\[
P = \sum_k P_k\,.
\]
The electron density $n$ (cf. Eq.~(\ref{eq:nexpr})) is given by
\[
n = \sum_k \mbox{diag}\left(\I P_k\Idag\right)\,.
\]
The electron-ion, exchange-correlation, and electron-Hartree energies
depend only on $n$, and provided the above $k$-dependent expression
for $n$ is used, these contributions require no further modification
from the forms already given in Eqs.~(\ref{eq:Eeiexpr}),
(\ref{eq:Excexpr}), and (\ref{eq:EeHexpr}) respectively.

The only change required to the basis-dependent operators involves the
use of the Laplacian for computing the kinetic energy.  The proper
generalization is to define $k$-dependent Laplacian matrices $L_k$
through
\[
(L_k)_{\alpha \beta} \equiv \int d^3r\ \left[e^{ik\cdot
r}b_\alpha(r)\right]^*\,\nabla^2\,\left[e^{ik\cdot
r}b_\beta(r)\right]\,.
\]
This immediately leads to the following expression for the kinetic
energy:
\[
T = -{1\over 2}\sum_k \mbox{Tr}\left(L_kP_k\right)\,.
\]
We still require the operator $L$ as defined in Eq.~(\ref{eq:Ldef})
for operations involving the Hartree field $\phi$ and the Poisson
equation.  The $L$ operator is $L_k$ evaluated at $k=0$.

The generalization of non-local potentials
(Appendix~\ref{appendix:nlpots}) to multiple $k$-points is also
straightforward.  The energy expression of Eq.~(\ref{eq:VPexpr})
generalizes to
\[
E_{nl} = \sum_k \mbox{Tr}(VP_k)\,.
\]

Having completed the specification of the Lagrangian with multiple
$k$-points, the generalizations required for the orthonormality
condition and the expressions for the derivatives of the Lagrangian
follow immediately.  We introduce overlap matrices $U_k$ and
unconstrained variables $Y_k$ (cf. Eqs.~(\ref{eq:Udef}) and
(\ref{eq:CYUmhalfVdag})),
\[
U_k = Y_k^\dag \calO Y_k\ \ \mbox{and}\ \ C_k = Y_k U_k^{-1/2}\,,
\]
where for simplicity we have set all subspace-rotation matrices to
identity, $V_k=I$.  The differential of the Lagrangian takes the form
(cf. Eq.~(\ref{eq:HdP}))
\[
d\LLDA  = \sum_k \mbox{Tr}(H_k\,dP_k)\,.
\]
The single-particle Hamiltonians $H_k$ depend on $k$ only through the
kinetic operators $L_k$,
\[
H_k = -{1\over 2}L_k + \Idag \left[\mbox{Diag
}V_{sp}\right]\I\ + V\,.
\]
The expression for the single-particle potential $V_{sp}$ is
unmodified from that of Eq.~(\ref{eq:Hspdef}) as it only depends on
the total electron density $n$.  The term $V$ is to be added only if
non-local potentials are employed (see
Appendix~\ref{appendix:nlpots}).

The expressions of Eq.~(\ref{eq:dLdYdag}) for the derivative of the
Lagrangian also generalize in the following straightforward way,
\begin{eqnarray*}
d\LLDA & = & \mbox{2 Re}\, \sum_k \mbox{Tr}\left[dY_k^\dag
\left({\partial \LLDA  \over \partial Y_k^\dag}\right)
\right],\ \mbox{where}\nonumber\\ \left({\partial \LLDA  \over
\partial Y_k^\dag}\right) & \equiv & w_k \left\{\left(I-
\calO C_kC_k^\dag\right)H_kC_kF_kU_k^{-1/2} +
\calO C_kQ_k([\tilde{H}_k,F_k])\right\},\ \mbox{and}\nonumber\\
\tilde{H}_k & \equiv & C_k^\dag H_kC_k\ ,
\end{eqnarray*}
where $Q_k$ is the natural generalization of the $Q$ operator
which uses the eigenvalues and eigenvectors of $U_k$
(Appendix~\ref{appendix:UmhalfQ}).

\section{Complete LDA code with $k$-points and non-local potentials}
\label{appendix:kptvnlcode}
In this section, we summarize and gather together the expressions for
the LDA Lagrangian and its derivatives in the DFT++ formalism for a
system with $k$-points and non-local potentials.  This type of system
provides the natural starting point for studying bulk systems and the
properties of defects in bulk-like systems \cite{RMP}.

As we have emphasized previously, it is sufficient for us to display
the formulae for the Lagrangian and its derivatives because formulae
in the DFT++ language specify all the operations that must be
performed and translate directly into computer code. (See
Section~\ref{sec:imploptpara}.)  Given the Lagrangian and its
derivatives, we can use a variety of methods to achieve
self-consistency. (See Section~\ref{sec:minalgs}.)

We follow the notation of Appendix~\ref{appendix:kpts} and refer the
reader to it for relevant details and definitions.  The point we wish
to emphasize is the compactness of the formalism and how it allows us
to specify an entire quantum-mechanical Lagrangian or energy function
in a few lines of algebra which explicitly show the operations
required for the computation.  We specialize to the case of
Kleinmann-Bylander \cite{KB} non-local potentials
(Appendix~\ref{appendix:nlpots}).
\begin{center}
\framebox[0.9\textwidth]{\parbox{0.9\textwidth}{
\begin{eqnarray*}
\LLDA & = & -{1\over 2} \sum_k w_k\,\mbox{Tr}\left(F_kC_k^\dag L_k
C_k\right) + \left(\J n\right)^\dag \left[V_{ion} + \calO\J\exc(n) -
\Obar\phi\right]\\ & & + \sum_k w_k \sum_s
\mbox{Tr}\left(\bar{M}_s(A_s^\dag C_k)F_k(A_s^\dag C_k)^\dag\right)
+{1\over 8\pi}\phi^\dag L\phi\,,\\ {\partial \LLDA \over \partial
Y_k^\dag} & = & w_k \left\{\left(I-\calO
C_kC_k^\dag\right)H_kC_kF_kU_k^{-1/2} + \calO C_kQ_k([C_k^\dag
H_kC_k,F_k])\right\}\,,\\ {\partial \LLDA \over \partial \phi^\dag} &
= & -{1\over 2}\Obar\J n + {1\over 8\pi}L\phi\,,\\ H_k & = & -{1\over
2}L_k + \Idag \left[\mbox{Diag }V_{sp}\right]\I + \sum_s
A_s\bar{M}_sA_s^\dag\,,\\ V_{sp} & = & \Jdag V_{ion} + \Jdag\calO\J
\exc(n) + [\mbox{Diag }\excprime(n)]\Jdag\calO\J n - \Jdag\Obar\phi\,.
\end{eqnarray*}}}
\end{center}

\section{The $Q$ operator}
\label{appendix:UmhalfQ}

In this appendix, we define the $Q$ operator which appears in
expressions for the derivative of the Lagrangian, e.g. in
Eq.~(\ref{eq:dLdYdag}).  The formal properties satisfied by $Q$ are
also presented, properties used in the derivation of the expression
for the derivative based on the connection between $Q$ and the
differential of the matrix $U^{1/2}$. (See the derivation starting
from Eq.~(\ref{eq:Hspdef}) and resulting in Eq.~(\ref{eq:dLdYdag}) in
Section~\ref{sec:Yderiv}.)

We start with the Hermitian matrix $U$.  Let $\mu$ be a diagonal matrix
with the eigenvalues of $U$ on its diagonal, and let $W$ be the
unitary matrix of eigenvectors of $U$. Thus, the following relations
hold: $U=W\mu W^\dag$, $W^\dag W = WW^\dag = I$, and $UW=W\mu$.

Consider the differential of the matrix $U$.  The Leibniz rule results
in
\[
dU = dW\mu\,W^\dag + W\,d\mu\,W^\dag + W\mu\,dW^\dag\,.
\]
Using the unitarity of $W$, we have
\[
W^\dag dU W = W^\dag dW \mu + \mu\,\, dW^\dag W + d\mu\,.
\]
Differentiating the relation $W^\dag W=I$, we have that $dW^\dag W +
W^\dag dW = 0$ or $dW^\dag W = - W^\dag dW$.  Substituting this above,
we arrive at the relation
\begin{equation}
W^\dag dU W = [W^\dag dW,\mu] + d\mu\,.
\label{eq:WdagdUW}
\end{equation}
This equation describes how differentials of the eigenvalues and
eigenvectors of $U$ are related to the differential of $U$, and it is
simply a convenient matrix-based expression of the results of
first-order perturbation theory familiar from elementary quantum
mechanics.  To see this equivalence, we first examine the diagonal
elements of Eq.~(\ref{eq:WdagdUW}) and find
\begin{equation}
d\mu_n = (W^\dag dU W)_{nn}\,,
\label{eq:dmun}
\end{equation}
the familiar expression for the first order shift of the eigenvalue
$\mu_n$.  Considering off diagonal matrix elements of
Eq.~(\ref{eq:WdagdUW}) leads to
\begin{equation}
(W^\dag dW)_{nm} = {(W^\dag dU W)_{nm} \over \mu_m - \mu_n} \ \ \
\label{eq:WdagdW}
\mbox{for } n \neq m\,,
\end{equation}
which is the expression for the first order shift of the $m$th
wave function projected on the $n$th unperturbed wave function.

Next, we consider $f(U)$, an arbitrary analytic function of $U$.
Using the eigenbasis of $U$, we can write $f(U) = Wf(\mu)W^\dag$
where by $f(\mu)$ we mean the diagonal matrix obtained by applying $f$
to each diagonal entry of $\mu$ separately.  Following the same logic
as above, the differential of $f(U)$ satisfies
\[
W^\dag d[f(U)] W = [W^\dag dW,f(\mu)] + f'(\mu)d\mu\,.
\]
Computing matrix elements of the above equation and using
Eqs.~(\ref{eq:dmun}) and (\ref{eq:WdagdW}), we arrive at the general
result
\begin{equation}
(W^\dag d[f(U)] W)_{nm} = (W^\dag dU W)_{nm} \cdot \left\{
\begin{array}{cc}
f'(\mu_n) & \mbox{if }m = n\\
\left[{f(\mu_m)-f(\mu_n) \over \mu_m-\mu_n}\right]
& \mbox{if }m \neq n
\end{array}
\right. .
\label{eq:dfU}
\end{equation}

We now apply this result to the case where $f(U)=U^{1/2}$.  This means
that $f(\mu_n)=\sqrt{\mu_n}$ and that $f'(\mu_n)=1/(2\sqrt{\mu_n})$ in
Eq.~(\ref{eq:dfU}).  By employing the algebraic identity
$(\sqrt{x}-\sqrt{y})/(x-y)=1/(\sqrt{x}+\sqrt{y})$, we arrive at the
expression
\begin{equation}
(W^\dag d[U^{1/2}] W)_{nm} = {(W^\dag dU W)_{nm} \over \sqrt{\mu_n} +
\sqrt{\mu_m}} = (W^\dag Q(dU) W)_{nm}\,,
\label{eq:UhalfQ}
\end{equation}
where we define the operator $Q(A)$ for an arbitrary matrix $A$ to be
\begin{equation}
(W^\dag Q(A) W)_{nm} \equiv {(W^\dag A W)_{nm} \over \sqrt{\mu_n} +
\sqrt{\mu_m}}\,.
\label{eq:Qdef}
\end{equation}
From this definition of $Q$, it is easy to prove that the following
identities are satisfied for arbitrary matrices $A$ and $B$ and
arbitrary power $p$:
\begin{eqnarray}
Q(dU) & = & d[U^{1/2}]\nonumber\\
\mbox{Tr}\left(Q(A)B\right) & = & \mbox{Tr}\left(AQ(B)\right)\nonumber\\
\mbox{Tr}\left(Q(U^pA)B\right) & = & \mbox{Tr}\left(U^pQ(A)B\right)\nonumber\\
\mbox{Tr}\left(Q(AU^p)B\right) & = & \mbox{Tr}\left(Q(A)U^pB\right)\nonumber\\
A & = & Q(A)U^{1/2} + U^{1/2}Q(A)\,.
\label{eq:Qidents}
\end{eqnarray}
These are the identities used in the derivation of the expression for
the derivative of the Lagrangian with respect to $Y$ in
Section~\ref{sec:Yderiv}.

\bibliographystyle{plain}

\begin{thebibliography}{99}

\bibitem{ACprl}{T.A. Arias, M.C. Payne, and J.D. Joannopoulos,
\emph{Physical Review Letters} {\bf 69}, 1077 (1992).}

\bibitem{KohnSham}{W. Kohn and L.J. Sham, \emph{Physical Review} {\bf
140}, A1133 (1965).}

\bibitem{fftw}{M. Frigo and S.G. Johnson, "FFTW: An Adaptive Software
Architecture for the FFT," Proc. ICASSP 1998, vol. 3, p. 1381.  The
software package is available at http://www.fftw.org}

\bibitem{LAE}{R. Lippert, T.A. Arias, and A. Edelman, \emph{Journal of
Computational Physics} {\bf 140} 270 (1998).}

\bibitem{wavelets}{T.A. Arias, {\em Reviews of Modern Physics} {\bf
71} 267 (1999).}

\bibitem{torkelarias}{T.A. Arias and T.D. Engeness, ``Beyond Wavelets:
Exactness theorems and algorithms for physical calculations'',
\underline{Computer Simulation Studies in Condensed Matter Physics
XII}, Eds. D. Landau, S.P. Lewis, and H.B. Schuttler (Springer Verlag,
Heidelberg, 1999).}

\bibitem{RMP}{M.C. Payne, M.P. Teter, D.C. Allan, T.A. Arias, and
J.D. Joannopoulos, \emph{Reviews of Modern Physics} {\bf 64}, 1045
(1992), and references therein.}

\bibitem{finiteel}{For recent developments, see J.E. Whiteman,
\underline{The Mathematics of Finite Elements and}\\
\underline{Applications: Highlights 1996} (John Wiley \& Sons, New
York, 1997); for an introduction, see T. Chandrapatha and
A. Belegundu, \underline{Introduction to Finite Elements in
Engineering} (Prentice Hall, Upper Saddle River, NJ, 1997).}

\bibitem{gauss}{E.R. Davidson and D. Feller, {\em Chemical Review}
{\bf 86} 681 (1986); S. Wilson, {\em Advances in Chemical Physics}
{\bf 67} 439 (1987).}

\bibitem{GunnarsonLundqvist}{O. Gunnarson and B.I. Lundqvist, {\em
Physical Review B}, {\bf 13}, 4274 (1976).}

\bibitem{ParrYang}{R.G. Parr and W. Yang,
\underline{Density-Functional Theory of Atoms and Molecules} (Oxford
U. Press, New York, 1989).}

\bibitem{PerdewZunger}{J. Perdew and A. Zunger, {\em Physical Review
B}, {\bf 23}, 5048 (1981).}

\bibitem{GoedeckerUmigar}{S. Goedecker and C.J. Umigar, {\em
Physical Review A}, {\bf 55}, 1765 (1997).}

\bibitem{Gonze}{X. Gonze, D.C. Allan, and M.P. Teter, {\em Physical
Review Letters} {\bf 68} 3603 (1992).}

\bibitem{Broyden}{D.D. Johnson, {\em Physical Review B} {\bf 38} 12807
(1988), and references therein.}

\bibitem{multigrid}{P. Wesseling, \underline{An Introduction to
Multigrid Methods} (Wiley, Chichester, New York, 1992).}

\bibitem{fastgauss}{S. Obara and A. Saika, {\em Journal of Chemical
Physics} {\bf 84} 3963 (1986).}

\bibitem{NumRecip}{W.H. Press, B.P. Flannery, S.A. Teukolsky, and
W.T. Vetterling, \underline{Numerical Recipes in C}, (Cambridge U.,
Cambridge, 1988).}

\bibitem{MVP}{N. Marzari, D. Vanderbilt, and M.C. Payne, {\em Physical
Review Letters} {\bf 79}, 1337 (1997).}

\bibitem{KB}{L. Kleinmann and D.M. Bylander, {\em Physical Review
Letters} {\bf 48}, 1425 (1982).}

\bibitem{fft}{J.W. Cooley and J.W. Tukey, {\em Mathematics of
Computation} {\bf 19} 297 (1965); P. Duhamel and M. Vetterli, {\em
Signal Processing} {\bf 19} 259 (1990).}

\bibitem{BCR}{G. Beylkin, R.R. Coifman, and V. Rokhlin, {\em
Commun. Pure and Appl. Math.} {\bf 44} 141 (1992).}

\bibitem{Sibars}{S. Ismail-Beigi and T.A. Arias, {\em Physical Review
B} {\bf 57} 11923 (1998).}

\bibitem{Sidisloc}{G. Csanyi, S. Ismail-Beigi, and T.A. Arias, {\em Physical
Review Letters} {\bf 80} 3984 (1998).}

\end{thebibliography}

\end{document}